\documentclass[12 pt]{JHEP3}
\usepackage[centertags]{amsmath}
\usepackage{amssymb}
\newcommand{\half}{{1 \over 2}}

\def\be{\begin{equation}}
\def\ee{\end{equation}}
\def\ba{\begin{array}{l}}
\def\ea{\end{array}}
\def\bea{\begin{eqnarray}}
\def\eea{\end{eqnarray}}
\def\beas{\begin{eqnarray*}}
\def\eeas{\end{eqnarray*}}

\date{}
\title {Counting Giant Gravitons in $AdS_3$}
\author{ Suvrat Raju$^{a,b}$ \\
\small{\emph{$^{a}$Department of Physics, Harvard University, Cambridge MA 02138, USA}}\\
\small{\emph{$^{b}$Department of Theoretical Physics,
                   Tata Institute of Fundamental Research,}}\\
\small{\emph{Homi Bhabha Road, Mumbai 400005, India}} }

\preprint{
    HUTP-07/A0003 \\
    TIFR/TH/07-17 \\
    \texttt{arXiv:0709.1171 [hep-th]}
}

\abstract{
We quantize the set of all quarter BPS brane probe solutions in global $AdS_3 \times S^3 \times T^4/K3$ found in \cite{Mandal}. We show that, generically, these solutions give rise to states in discrete representations of the $SL(2,R)$ WZW model on $AdS_3$.
Our procedure provides us with a detailed description of the low energy ${1 \over 4}$ and ${1 \over 2}$ BPS sectors of string theory on this background. The ${1 \over 4}$ BPS partition function jumps as we move off the point in moduli space where the bulk theta angle and NS-NS fields vanish. 
We show that generic ${1 \over 2}$ BPS states are protected because
they correspond to geodesics rather than puffed up branes. By exactly quantizing the simplest of the probes above, we verify 
our description of ${1 \over 4}$ BPS states and find 
agreement with the known spectrum of ${1 \over 2}$ BPS states of 
the boundary theory. 
We also consider the contribution of these probes to the elliptic genus and discuss puzzles, and their possible resolutions, in reproducing the elliptic genus of the symmetric product. 
}

\begin{document}


\section{Introduction}
Recently, Mandal, Raju and Smedb\"ack constructed the moduli space of all quarter BPS brane probes in Type IIB string theory on 
global $AdS_3 \times S^3 \times T^4/K3$ \cite{Mandal}.  These brane probes consist of D1 branes, D5 branes that wrap the internal manifold and bound states of D1-D5 branes. The condition for supersymmetry translates, roughly, into the statement that these branes must always be `left-moving'. In this paper, we will attempt to quantize this set of solutions. This provides a complete description of the low energy structure of the ${1 \over 4}$ BPS sector of string theory in this background.

The AdS/CFT conjecture\cite{Maldacena:1997re} relates type IIB superstring theory on global $AdS_3 \times S^3 \times T^4/K3$  to the NS sector of a (4,4) CFT living on the boundary of AdS. The NS-sector of the $N=4$ algebra in 1+1 dimensions has short representations that are built on a special kind of lowest weight state called a chiral primary (A chiral primary has the property that its R-charge is equal to its conformal weight).  ${1 \over 4}$ BPS states of the boundary theory are of the form $|{\rm anything} \rangle |{\rm chiral~primary} \rangle$. The probe solutions that we will discuss are dual to these states.

One of the exciting features of these probes in global AdS is that, for a generic assignment of charges (i.e spacetime momenta), 
they are bound to the center of AdS and cannot escape to infinity. This, however, makes the quantization of these probes difficult since, in the interior of AdS,  the natural symplectic structure on the space of solutions couples different degrees of freedom to each other in a complicated manner. 
To circumvent this difficulty, we first rewrite the supersymmetric probe solutions of \cite{Mandal} as left-moving classical solutions of a `Polyakov' type non-linear sigma model. The `bound states' above then give rise to states in discrete representations of the $SL(2,R)$ WZW model on $AdS_3$.  

The supersymmetric solutions we have described have the property that they exist only when the bulk theta angle and NS-NS fields are set to zero. 
Now, it is well known that on this special submanifold of moduli space the boundary theory is singular because the stack of D1 and D5 branes that make up the background can separate at no cost in energy \cite{Seiberg:1999xz}. 
This leads to the presence of a continuum in the spectrum that vanishes as soon as we turn on a theta angle or NS-NS fields. It is natural to ask if this continuum meets the space of ${1 \over 4}$ BPS states. Generically, as we have explained, ${1 \over 4}$ BPS states are described by discrete states that do not lie in a continuum. However, it turns out, that for a very special assignment of charges, supersymmetric probes in AdS can escape to infinity. Semi-classically, the quantization of these special probes gives rise to states at the bottom of continua.

So, when the theta angle and NS-NS fields are set to zero, the ${1 \over 4}$ BPS partition function has an intricate structure, that we will describe in some detail, with unambiguous contributions from all the discrete states.  As soon as we turn on one of the bulk moduli above, the ${1 \over 4}$ BPS partition function jumps. Such a process can happen when short representations combine in pairs to form long representations. For example, in ${\cal N} = 4$ Yang Mills theory on $S^3 \times R$, which is dual to type IIB string theory on $AdS_5 \times S^5$ it is known that both the ${1 \over 16}$ and ${1 \over 8}$ BPS partition functions jump as soon as we turn on a 't Hooft coupling and are not further
renormalized \cite{Kinney:2005ej}. 

However, by taking appropriate limits of the ${1 \over 4}$ BPS partition function, one may obtain two `protected' quantities: the elliptic genus and
 the spectrum of ${1 \over 2}$ BPS states (states that are built on a lowest weight state of the form $ |{\rm chiral~primary} \rangle|{\rm chiral~primary} \rangle $).\begin{footnote}{Since the terms ${1 \over 4}$ BPS partition function and elliptic genus are, unfortunately, sometimes used interchangeably in the literature, we pause here to review our terminology. In a (4,4) theory, states may be indexed by their left and right moving conformal weights $h, \bar{h}$ and R-charges $r,\bar{r}$. The partition function depends on 4 chemical potentials:
\begin{equation}
Z(\beta, \bar{\beta}, \rho, \bar{\rho}) = {\rm Tr} e^{-\beta h - \bar{\beta} \bar{h} - \rho r - \bar{\rho} \bar{r}} . 
\end{equation}
The ${1 \over 4}$ BPS partition function depends on 3 chemical potentials and is given by:
\begin{equation}
Z_{1 \over 4}(\beta, \rho, \bar{\mu}) = \lim_{\bar{\beta} \rightarrow \infty} Z(\beta, \bar{\beta}, \rho, -\bar{\beta} + \bar{\mu}) .
\end{equation}
The elliptic genus depends on 2 chemical potentials and is given by:
\begin{equation}
E(\beta, \rho) = Z(\beta, \bar{\beta}, \rho, -\bar{\beta} + 2 \pi i),
\end{equation}
where the RHS is actually independent of $\bar{\beta}$. 
The ${1 \over 2}$ BPS partition function also depends on 2 chemical potentials:
\begin{equation}
\label{halfbpsdef}
Z_{1 \over 2}(\mu,\bar{\mu}) = \lim_{\beta \rightarrow \infty, -\bar{\beta} \rightarrow \infty} Z(\beta, \bar{\beta}, -\beta + \mu, -\bar{\beta}+ \bar{\mu})  .
\end{equation}
}
\end{footnote} These quantities are protected in that they do not change as we move about
on the moduli space unless the spectrum changes discontinuously at some point.
 So it is of interest to compare our results for these quantities with 
their known values at the point in moduli space where the boundary theory becomes a symmetric product.

Now, de Boer, building on \cite{Maldacena:1998bw}, found that the low energy structure of the ${1 \over 2}$ BPS partition function and elliptic genus of the symmetric product had a striking property \cite{deBoer:1998us,deBoer:1998ip}. He found that these partition functions, for energies lower than the BTZ black hole threshold, were completely explained by gravitons and multi-gravitons, with an appropriate exclusion principle.\begin{footnote}{This range of energies is what contributes to the `polar-part' of the elliptic genus. Since the elliptic genus can be almost completely reconstructed from a knowledge of its polar part \cite{Dijkgraaf:2000fq}
 it appears that this index in $AdS_3$ knows only about gravitons, just like its counterpart in $AdS_5$ \cite{Kinney:2005ej}. }\end{footnote} The discussion above gives us, for the first time, a clear explanation of this phenomenon. Supersymmetric giant graviton solutions do not exist at a generic point in moduli 
space. Hence, almost everywhere in moduli space, the ${1 \over 2}$ BPS partition function and elliptic genus are protected and see contributions only from gravitons and multi-gravitons at low energies.

However, we are also left with a puzzle because on this special submanifold of moduli space, giant graviton solutions {\em do} exist far below the black-hole threshold. Why is their signature not seen in the ${1 \over 2}$ BPS partition function and elliptic genus evaluated at the symmetric product point?
In section \ref{chiralchiralsection}, we resolve half of this puzzle by showing that, except at very special charges, giant gravitons cannot describe ${1 \over 2}$ BPS states! The classical solutions that correspond to generic ${1 \over 2}$ BPS states, all the way up to the threshold of the BTZ black hole, are geodesics and not puffed up branes -- gravitons rather than giant gravitons. The question of the elliptic genus is more subtle. The elliptic genus is `blind' to right-moving charges. So, semi-classically, the sum that contributes to the elliptic genus runs not only over `bound states' but also over the states at the bottom of continua. The presence of this continuum removes the puzzle since it invalidates the usual arguments that protect the index.

To verify this semi-classical story, we exactly quantize the simplest of the probes above -- the D-string -- by dualizing to an F1-NS5 frame and using the techniques of \cite{Maldacena:2000hw,Maldacena:2000kv}.  This analysis yields results that are almost entirely in accordance with our semi-classical expectations. We view this as a validation of our basic philosophy that the supersymmetric sector of the full quantum theory may be understood by quantizing supersymmetric classical solutions (this idea has previously been exploited in \cite{Grant:2005qc,Mandal:2005wv,Maoz:2005nk,Rychkov:2005ji,Biswas:2006tj,Mandal:2006tk,Martelli:2006vh,Basu:2006id}). We find, as expected, discrete ${1 \over 4}$ BPS states that, moreover, obey exactly the same energy formula that we obtain by semi-classical methods. By taking limits of the ${1 \over 4}$ BPS partition function, we are also able to reproduce, almost exactly, the spectrum of ${1 \over 2}$ BPS states of the symmetric product. However, we find, as has been found earlier \cite{Argurio:2000tb} and as is expected from an analysis of the singularities of the boundary theory on this submanifold of moduli space \cite{Seiberg:1999xz}, that some chiral-primaries are missing. These missing chiral-primaries are exactly at the point where, semi-classically, we expect to find a continuum. However, in the exact 
analysis, the measure for the continuum vanishes at this point. 
We discuss this issue further in Section \ref{spacetimepartfun}.

A brief outline of this paper is as follows. In section \ref{review}, we review the construction of ${1 \over 4}$ BPS brane probe solutions described in \cite{Mandal} and discuss some toy examples of semi-classical quantization. In section \ref{polyasection}, we describe a second approach to classical supersymmetric solutions that turns out to be much more convenient for purposes of quantization. In section \ref{semiclassicalsection}, we discuss the quantization of these probe solutions and show that they correspond to states in discrete representations of the $SL(2,R)$ WZW model. We also describe the resultant Hilbert space in the semi-classical approximation. In Section \ref{spacetimepartfun}, we perform an independent and exact quantization of the simplest of the probes above -- D strings. By restricting the partition function of D-strings to its supersymmetric subsector, we validate the energy formula of section \ref{semiclassicalsection}. 
We also use this exact calculation to discuss, more precisely, the contribution of these probes to the elliptic genus and the half-BPS partition function; the 
latter matches very well with the result expected from the symmetric product.
\section{Classical Supersymmetric Solutions in Global $AdS_3$} 
\label{review}
\subsection{Review}
We start this section with a self-contained review of the relevant parts of \cite{Mandal}, although we present only the results and no proofs.

Consider global $AdS_3 \times S^3 \times T^4$ with metric:
\begin{equation}
\label{globaladsmetric}
\begin{split}
&ds^2= G_{\mu \nu} d x^{\mu} d x^{\nu} \\
&= g \sqrt{{Q_1 Q_5 \over v}} \alpha' \left[-\cosh^2{\rho} dt^2 + \sinh^2{\rho} d \theta^2  + d \rho^2 + d \zeta^2 + \cos^2 \zeta d \phi_1^2 + \sin^2 \zeta d \phi_2^2\right]\\ &+ \sqrt{Q_1 \over Q_5 v} \alpha' d s^2_{\rm int} .
\end{split}
\end{equation}
$ds_{\rm int}^2$ is the metric on the internal $T^4$ whose sides are of length $2 \pi v^{1 \over 4}$ and  $g, Q_1, Q_5$ are parameters that determine the string coupling constant, and the 3 form  and 7 form RR field strengths according to the formulae summarized in Table \ref{global} below. We are following the notation of \cite{Maldacena:1998bw}. 
We parameterize the internal manifold using the coordinate $z^{1 \ldots 4}$. Although we will concentrate here on the case where the internal manifold is $T^4$, our results may be easily generalized to $K3$. 

If the theta angle (a linear combination of the RR 0 and 4 form) and NS-NS fields are set to zero then, as we explain below, this background supports ${1 \over 4}$ BPS brane probes that consists of D1 branes, D5 branes and also bound states of $p$ D1 and $q$ D5 branes.

First, consider the case of a D-string (i.e $p = 1, q = 0$). The bosonic part of the brane action is:
\begin{equation}
\label{probebrane} 
S=\int {\cal L}_{\rm brane} \, d \tau d \sigma = -{1 \over 2 \pi \alpha'} \int
e^{-\phi} \sqrt{-h} \, d \tau d \sigma + {1 \over 2 \pi \alpha'} \int B_{\mu \nu}
\partial_\alpha X^\mu \partial_\beta X^\nu {\epsilon^{\alpha \beta} \over 2}  \, d \tau d \sigma  .
\end{equation}
where $h = \det(G_{\mu \nu} \partial_{\alpha} X^{\mu} \partial_{\beta} X^{\nu})$, $\alpha, \beta$ run over the two worldsheet coordinates $\sigma, \tau$ and $B$ is the RR 2 form and $\phi$ the dilaton specified in Table \ref{global}. Now, consider the `effective' metric:
\begin{equation}
\label{effectivemetric}
\begin{split}
&ds^2= G^{\rm eff}_{\mu \nu} d x^{\mu} d x^{\nu} \\
&= Q_5  \left[-\cosh^2{\rho} dt^2 + \sinh^2{\rho} d \theta^2  + d \rho^2 + d \zeta^2 + \cos^2 \zeta d \phi_1^2 + \sin^2 \zeta d \phi_2^2\right] + {1 \over g} d s^2_{\rm int} .
\end{split}
\end{equation}
If we define $h^{\rm eff} = \det\left(G^{\rm eff}_{\mu \nu} \partial_{\alpha} X^{\mu} \partial_{\beta} X^{\nu}\right)$ then {\em classically} the action \eqref{probebrane} may be rewritten as:
\begin{equation}
\label{probebraneeff}
S=\int {\cal L}_{\rm brane} = -{1 \over 2 \pi } \int
\sqrt{-h^{\rm eff}} \, d \tau d \sigma + {1 \over 2 \pi \alpha'} \int B_{\mu \nu}
\partial_\alpha X^\mu \partial_\beta X^\nu {\epsilon^{\alpha \beta}
\over 2} \, d \tau d \sigma .
\end{equation}
$Q_1$ does not appear in the effective action above and this explains why, only $Q_5$ and not $Q_1$ appears in the formulae of Table \ref{global}. 

If we denote the energy and angular momenta in global $AdS$ by $E,L$ respectively and the two $SU(2)$ angular momenta on the $S^3$ by ${J_1 + J_2 \over 2}, {J_1 - J_2 \over 2}$ then the bulk BPS bound is
\begin{equation}
\label{bpsbound}
E - L \geq J_1 + J_2.
\end{equation}
It was found in \cite{Mandal} that probe D-strings saturate this bound provided the vector $n^{\mu} = {\partial \over \partial t} + {\partial \over \partial \theta} + {\partial \over \partial \phi_1} + {\partial \over \partial \phi_2}$ is tangent to the brane worldvolume at all points.

Now, let us say we are given the shape of the D-string at a particular point of time. We can then translate each point on the string along the integral curves of the null vector field above to generate the entire brane worldvolume. Hence, the set of all supersymmetric brane worldvolumes is the same as the set of all initial shapes of the D-string. 

The brane worldvolume is parameterized by 10 functions $X^{\mu}(\sigma, \tau)$. Let us now choose the coordinate $\tau$ along the brane worldvolume to be such that 
\begin{equation}
{\partial X^{\mu} \over \partial \tau} = n^{\mu}.
\end{equation}
In the coordinate system of \eqref{globaladsmetric}, $n^{\mu}$ is just a constant so we can explicitly solve the equation above for the functions $X^{\mu}$. We find that:
\begin{equation}
\label{explicitsolutions}
t=\tau,~\theta=\theta(\sigma)+\tau,~\rho=\rho(\sigma),~\zeta=\zeta(\sigma),~\phi_1=\phi_1(\sigma)+\tau,~\phi_2=\phi_2(\sigma)+\tau,~z^a = z^{a}(\sigma) .
\end{equation} 
Hence, the set of all supersymmetric D-strings may be parameterized (up to a reparameterization of $\sigma$) by the set of all profile functions $\theta(\sigma), \rho(\sigma), \phi_1(\sigma), \phi_2(\sigma), z^a(\sigma)$.\begin{footnote}{In order for the brane worldvolume to satisfy the equations of motion, it is important that the determinant of the worldsheet metric not vanish at any point. In the parameterization above, this means $X' \cdot \dot{X}$ must maintain a constant sign}\end{footnote} In table \ref{global} we summarize these results and also evaluate the spacetime momenta (that integrate to give conserved charges of the action \eqref{probebraneeff}) on these solutions.
\TABLE[!h]{\caption{D branes in Global AdS}
\label{global}
\begin{tabular}{|l|}
\hline
{\bf Geometry} \\
${ds^2 \over \alpha'} = l^2 \left[-\cosh^2{\rho} dt^2 + \sinh^2{\rho} d \theta^2  + d \rho^2 + d \zeta^2 + \cos^2 \zeta d \phi_1^2 + \sin^2 \zeta d \phi_2^2\right] + \sqrt{Q_1 \over Q_5 v} d s^2_{\rm int}$ \\
$e^{-2 \phi} = {Q_5 v \over g^2 Q_1}, l^2 = {g \over \sqrt{v}} \sqrt{Q_1 Q_5}$ \\
${H_3 \over \alpha'} = {*H_7 \over \alpha'} = {d B \over \alpha'} =  Q_5 \sin{2 \zeta} d \zeta \wedge d \phi_1 \wedge d \phi_2 +  Q_5 \sinh(2 \rho) d \rho \wedge d t \wedge d \theta$ \\
${B \over \alpha'}=-{Q_5 \over 2 } \left[\cos 2 \zeta d \phi_1 \wedge d
\phi_2 - (\cosh(2 \rho) - 1) d t \wedge d \theta \right]$ \\
{\bf BPS Condition}\\
$E - L - J_1 - J_2 = -\int(P_t + P_{\theta} +
\tilde P_{\phi_1} +\tilde  P_{\phi_2})\, d \sigma = 0$\\
{\bf Null Vector tangent to worldvolume:} \\
$n^{\mu} = {\partial \over \partial t} + {\partial \over \partial \theta} + {\partial \over \partial \phi_1} + {\partial \over \partial \phi_2}$\\
{\bf Solution}\\
$t=\tau,~~\theta=\theta(\sigma)+\tau,~~\rho=\rho(\sigma)$ \\
$\zeta=\zeta(\sigma),~~\phi_1=\phi_1(\sigma)+\tau,~~\phi_2=\phi_2(\sigma)+\tau$ \\
$z^a_{\rm int} = z^{a}_{\rm int}(\sigma)$ \\
{\bf Momenta:}\\
$\gamma = {\sinh^2 \rho \theta^{'2} + \cos^2 \zeta \phi_1^{'2} +\sin^2 \zeta \phi_2^{'2} + \zeta'^2 + \rho^{'2} + {g^{\rm int}_{ab} \over Q_5 g} z^{a'} z^{b'} \over  \cos^2 \zeta \phi_1'+ \sin^2 \zeta \phi_2'+ \sinh^2 \rho \theta'}$ \\
$P_{t} ={Q_5 \over 2 \pi} \left[ -\gamma \cosh^2
\rho + \sinh^2{\rho} \theta' \right]$  \\
$P_\theta = {-Q_5 \over 2 \pi} \left[ \left(-\gamma + \theta' \right) \sinh^2 \rho \right]$ \\  
$\tilde P_{\phi_1}={-Q_5 \over 2 \pi} \left[  \left(-\gamma  + \phi_1' \right) \cos^2 \zeta + 
\half \left( \cos 2 \zeta -1 \right) \phi_2' \right] $\\ 
$\tilde P_{\phi_2} ={-Q_5 \over 2 \pi} \left[ \left( -\gamma + \phi_2' \right) \sin^2 \zeta - \half \left( \cos 2 \zeta + 1 \right) \phi_1' \right]$ \\
$P_{\rho} = {-Q_5 \over 2 \pi} \rho'$\\
$P_{\zeta}= {-Q_5 \over 2 \pi} \zeta'$ \\
$P_{z^{a}} = {-1 \over 2 \pi g}\left[g^{int}_{a b} z^{b'}\right] ~~ {\rm (internal~manifold)}$\\
\hline
\end{tabular}}

Now we turn to D5 branes. It may be shown, either by a kappa symmetry analysis or an analysis of the DBI action, that D5 branes that wrap the internal manifold and have the property that the vector $n^{\mu}$ is tangent to their worldvolume at each point are also supersymmetric \cite{Mandal}. The formulae for the momenta in Table \ref{global} are then all valid but with $Q_5$ replaced by $Q_1$. 

The third and last kind of supersymmetric probe is a bound state of $p$ D1 branes and $q$ D5 branes. To obtain a supersymmetric probe of this kind, we start with a stack of  coincident $q$ D5 branes all of which maintain the killing vector $n^{\mu}$ tangent to their worldvolume at each point. Now, we turn on $U(q)$ gauge fields on the worldvolume: $A_{{i}}(\sigma)$. These are translationally invariant along $\tau$ and give rise to a field strength:
\begin{equation}
F=  F_{\sigma i} d \sigma \wedge d z^{i} + {1 \over 2} F_{i j} d z^{i} \wedge d z^{j}.
\end{equation}
The condition for supersymmetry then is that this field strength be self-dual on the internal manifold: $F_{i j} = \epsilon_{i j}^{k l} F_{k l}$. We interpret this configuration as being a supersymmetric bound state of $q$ D5 branes and $p$ D1 branes, where $p$ is the instanton number of $F$
\begin{equation}
p = {1 \over 8 \pi^2} \int_{M_{\rm int}} {\rm Tr} (F \wedge F),
\end{equation}
and $F$ is normalized in the conventional way. 
These classical instanton configurations have moduli and instead of using the gauge fields $A_{i}(\sigma)$ it is convenient to parameterize them in terms of their moduli ${\cal \zeta}^a(\sigma)$. Note that the moduli can vary as a function of $\sigma$ without spoiling supersymmetry.

Somewhat surprisingly, the formulae for the momenta presented in Table \ref{global} continue to be valid for such $(p,q)$ strings with the following two generalizations:
\begin{enumerate}
\item
$Q_5$ is replaced by \begin{equation}\label{kdef}k = p (Q_5 - q) + q (Q_1 - p). \end{equation}
\item
the internal manifold $M_{\rm int}$ is replaced by the  moduli space of p instantons in a $U(q)$ theory on $M_{\rm int}$. We will denote this manifold by ${\cal M}_{p,q}$. For uniformity of notation, we will henceforth use ${\cal M}_{1,0} \equiv M_{int}$. The coordinates $z^{a}$ will also be used for ${\cal M}_{p,q}$.
\end{enumerate}
This result relies on the fact that classically, within the DBI approximation, the dynamics of the supersymmetric subsector of the 5+1 dimensional D5 brane theory reduces to the 
dynamics of a  1+1 dimensional sigma-model, without taking taking an IR limit!

The probe solutions listed above have several salient features
\begin{enumerate}
\item
They have an energy gap -- $E \geq \min\{Q_5,Q_1\}$. 
This is intuitive because below this energy one would expect the Hilbert space to comprise gravitons. At the minimum energy above, stringy effects in the form of these supersymmetric giant gravitons make their appearance.
\item
In the AdS background that we have been discussing, we can turn on self dual NS-NS fluxes on the internal manifold and a theta angle. On the boundary, this corresponds to deforming the theory with some marginal operators \cite{Dijkgraaf:1998gf}.  The formulae of Table \ref{global} are valid on the submanifold of moduli space where the coefficients of these operators are set to zero. If we move off this submanifold, there are no BPS giant graviton solutions. This means that the ${1 \over 4}$ BPS partition function which, as we will find, has an intricate structure on this submanifold jumps as soon as move off it. The only ${1 \over 4}$ BPS states at a generic point in moduli space are then given by the ${1 \over 2}$ BPS gravitons and multi-particles of these. This explains why the low energy elliptic genus and ${1 \over 2}$ BPS partition function of the symmetric
product do not see contributions from the ${1 \over 4}$ BPS giant gravitons that we have described.
\item
Now it is well known that on the special submanifold of moduli space that we have been considering, the boundary theory is singular \cite{Seiberg:1999xz}.
This raises the question as to whether the states obtained by quantizing the solutions of Table \ref{global} are somehow localized about the singularities of the Higgs branch. In particular, one may worry about whether these states are located in a continuum. That this is not so, can be seen from the fact that for generic charges, these solutions are bound to the interior of AdS and cannot go off to the boundary of AdS.

Consider, a  D-string near the boundary of $AdS$. Such a string can have finite energy only if the flux through the string almost cancels its tension. Hence, it must wrap the $\theta$ direction and we can use our freedom to redefine
 $\sigma$ to set $\theta'=w$. For such a string, if we take the strict $\rho \rightarrow \infty$ limit, we obtain
\begin{equation}
\begin{split}
&E - L =  {Q_5 \over 2 \pi}\int \gamma d \sigma \\
 &= {Q_5 \over 2 \pi} \int  \left[{\sinh^2 \rho \theta^{'2} + \cos^2 \zeta \phi_1^{'2} +\sin^2 \zeta \phi_2^{'2} + \rho^{'2} + G_{ab}X^{a'} X^{b'} \over  \cos^2 \zeta \phi_1'+ \sin^2 \zeta \phi_2'+ \sinh^2 \rho \theta'}\right] d \sigma \\
&= {Q_5 w} . 
\end{split}
\end{equation}
Thus, we notice that for strings stretched close to the boundary, the quantity $E-L$ must be quantized in units of $Q_5$. For intermediate, and generic, values of $E-L$ the solutions of Table \ref{global} are `bound' to the center of $AdS$. This indicates that quantizing them would lead to discrete states, rather than states that are at the bottom of a continuum.
\end{enumerate}

Let us elucidate point (3) above by considering another subset of solutions that do not wrap the $\theta$ circle at all. Consider the following solution (parameterized by $w,~\rho_0,~\zeta_0,~\phi_{1_0},~\theta_0$) 
\begin{equation}
\label{toymodelone}
t = \tau,~~\theta(\sigma)=\theta_0,~~\rho(\sigma)=\rho_0,~~\zeta(\sigma)=
\zeta_0,~~\phi_1(\sigma)=\phi_{1_0} + w \sigma,~~\phi_2(\sigma)=w \sigma . 
\end{equation}
Note, that we can absorb the constant in $\phi_2(\sigma)$ into a shift in the origin of $\sigma$.
For this solution (using $w>0$ which is necessary for supersymmetry)
\begin{equation}
\label{toyonemomenta}
E = Q_5 w \cosh^2(\rho_0),~~L = Q_5 w \sinh^2(\rho_0),~~ 
J_{1}=Q_5 w \sin^2(\zeta_0),~~J_{2} = Q_5 w \cos^2(\zeta_0) . 
\end{equation}
In this subsector, a given set of charges fixes $\rho_0$:
\begin{equation}
\sinh^2{\rho_0} = {L \over w Q_5} . 
\label{size-bound-q1}
\end{equation}

Equation \eqref{size-bound-q1} has a resemblance to the formula for the size of the extremal BTZ black-hole and we refer the interested reader to \cite{Mandal} for the details of this analogy. 
This discussion provides us with an inkling of one of the main results of this paper. Quantizing classical supersymmetric solutions in global $AdS$ generically leads to `bound' states. 

\subsection{Quantization using the DBI Action: Preliminary Attempts}
\label{prelimsemiclassical}
The space of all classical solutions of a theory is isomorphic to its phase space. The Lagrangian equips this space with a symplectic structure.  This may be used to canonically quantize the theory. The advantage of this approach is that it maintains covariance. Furthermore, we can restrict attention to a subsector of phase space by identifying the corresponding classical solutions. This technique was, it seems, invented by Dedecker \cite{dedecker1953cvf}, studied in  \cite{goldschmidt1973hcf,Kijowski:1973gi,gawedzki1974cfl,szczyrba1976sss,garcia570rss} and later brought back into use by \cite{zuckerman1987apa,crnkovic1987cdc}. We refer the reader to \cite{Lee:1990nz} for a nice exposition of this method.  

The philosophy of this paper is that it may be possible to quantize special subsectors of solutions, for example supersymmetric subsectors, to obtain a subset of the full Hilbert space. We have enumerated all low energy supersymmetric classical solutions to Type IIB string theory on global $AdS_3$ in the previous subsection. Unfortunately it is not technically feasible to quantize all these solutions using the action \eqref{probebraneeff} and its associated symplectic form. In section \ref{semiclassicalsection}, we will show how this problem may be attacked using another method. For this subsection, however, we will restrict attention to even smaller subsectors. There is no strict justification for this since the symplectic form does couple the subset of solutions we will discuss below to other solutions not in this subset. Yet, these studies are useful as {\em toy examples} that yields some insight into the structure of the quantum theory. 

To start with let us consider the subset of solutions \eqref{explicitsolutions} where we restrict to: 
\begin{equation}
\label{phiwinding}
\theta(\sigma)=0,~~\rho(\sigma) = \zeta(\sigma) = 0,~~\phi_1(\sigma) = \phi_2(\sigma) = w \sigma .
\end{equation}
with an arbitrary profile on the internal manifold ${\cal M}_{p,q}$.  For large $p,q$ this is not a severe restriction since most of the degrees of freedom of the string are in the fluctuations on ${\cal M}_{p,q}$. 

The profile of the string on the classical instanton moduli space ${\cal M}_{p,q}$ is parameterized by functions $z^a(\sigma)$, with conjugate momenta $P_{z^a} = -{1 \over 2 \pi g} g_{a b}^{\rm int} (z^b)'$ where $g^{\rm int}_{a b}$ is the metric on ${\cal M}_{p,q}$. The spacetime energy and angular momentum are given by:
\begin{equation}
\label{semiclassicaldirectwo}
{E+L \over 2} = {k w \over 2}  + {h_{\rm int} \over w},~~~~{E - L \over 2} = {k w \over 2} \, .
\end{equation}
where $h_{\rm int} = {1 \over 2 \pi g} \int g^{\rm int}_{a b} (z^{a})' (z^{b})' d \sigma $ is the `level' of the sigma model on ${\cal M}_{p,q}$. 

To see what happens when we quantize the canonical structure above, consider the space of functions $X(\sigma)$ with the symplectic form:
\begin{equation}
\label{symplecticscalar}
\Omega = \int -\delta X'(\sigma) \wedge \delta X(\sigma) {d \sigma \over 2 \pi} .
\end{equation}
Expanding $X(\sigma) = {X_{n} \over \sqrt{2 |n|}} e^{i n \sigma}$, the symplectic structure \eqref{symplecticscalar} leads to the usual Dirac bracket prescription:
\begin{equation}
\{X_{n}, X_{-n}\}_{\rm D.B}  = i, ~~~ n > 0
\end{equation}
Promoting Dirac brackets to commutators will lead to a Fock space that has the usual left moving oscillator modes of a scalar field, but no right 
moving oscillators or momentum zero modes. Since these zero modes are what tie the left and right movers together, what we have here is the purely `left-moving' part of a scalar field.

In exactly the same way, in the example above, we obtain the left-moving part of the quantum non-linear sigma model on ${\cal M}_{p,q}$. We will denote this Hilbert space, that comprises the trivial zero mode sector, by $H^0({\cal M}_{p,q})$. The energy in $AdS$ is related to the level of this CFT by the formula \eqref{semiclassicaldirectwo}. 

The sigma-model on ${\cal M}_{p,q}$ is conformal and admits an $N=4$ supersymmetric extension. Now, the left moving level of the boundary theory is given by ${E + L \over 2}$ and one may think that the superconformal algebra carries over from the worldsheet to spacetime via formula \eqref{semiclassicaldirectwo}. The usual Virasoro algebra (see \eqref{sugawaralgebra}) is indeed invariant under the redefinition $L'_{n}-\delta_{n,0} {c' \over 24} = {1 \over w}(L_{w n} - \delta_{n,0} {c \over 24}), c' = c w$, but now we see that the shift ${k w \over 2}$ does not allow us to use this prescription for equation \eqref{semiclassicaldirectwo}. In Section \ref{semiclassicalsection}, we will see how the $N=4$ sigma model on ${\cal M}_{p,q}$ is supplemented with degrees of freedom from the `center of mass' coordinates that shift the central charge to correctly generate this shift. 

To make the example above technically tractable, we were forced to fix the `center of mass' coordinates of the strings. We will, now,  relax this assumption slightly and consider the slightly different subset of solutions where we fix to 
\begin{equation}
\label{thetawinding}
\theta = w \sigma + \tau,~~\phi_1 = (\phi_1)_0 + \tau,~~ \phi_2 = (\phi_2)_0 + \tau.
\end{equation}
where $(\phi_1)_0, (\phi_2)_0$ are two real constants and $w$ an integer. $\rho, \zeta$ and the profile on ${\cal M}_{p,q}$ remain arbitrary. On this submanifold, we can expand out the $\rho$ and $\zeta$ in \eqref{explicitsolutions} as:
\begin{equation}
\rho(\sigma)= \sum_{-\infty}^{\infty} {\rho_n \over \sqrt{2 k |n|}} e^{i n \sigma},~~ \zeta=\sum_{-\infty}^{\infty} {\zeta_n \over \sqrt{2 k |n|}} e^{i n \sigma} . 
\end{equation}
The momenta of Table \ref{global}, then lead to the 
Dirac bracket prescriptions, for $n > 0$:
\begin{equation}
\label{diracprescriptions}
\begin{split}
-i \{\rho_{n}, \rho_{-n}\}_{\rm D.B} &= 1,  \\
-i \{\zeta_{n}, \zeta_{-n}\}_{\rm D.B} &= 1 . 
\end{split}
\end{equation}
Promoting these Dirac brackets to commutators leads, as we explained above, to the left-moving sector of the Hilbert space of a free scalar field.
Already, we see that the spacetime momenta do not have simple quadratic expressions in terms of the `creation' and `annihilation' operators above, except for 
\begin{equation}
\label{directsemiclassical}
L = {N_{\rho} + N_{\zeta} + h_{\rm int} \over w} . 
\end{equation}
where $N_{\rho}, N_{\zeta}$ are the levels of the $\rho$ and $\zeta$ CFT and $h_{\rm int}$ is the level of the non-linear sigma model on ${\cal M}_{p,q}$.\begin{footnote}{It is known that in the full quantum theory, both the $\rho$ CFT and the $\zeta$ CFT develop linear dilaton terms but we cannot derive these shifts in the stress tensor from our semi-classical perspective.}\end{footnote} 

The two examples above give us some insight into the structure of the full quantum theory. For example, we see that the excitations on the internal manifold ${\cal M}_{p,q}$ enter the formulae for spacetime energy and angular momentum in the simple fashion specified by \eqref{directsemiclassical} and \eqref{semiclassicaldirectwo}. We will obtain similar formulae in the full quantization that we perform in section \ref{semiclassicalsection}.  

Unfortunately, it does not seem technically possible to proceed and quantize the entire space of solutions in Table \ref{global} by extending these techniques. So, we will turn, in the next section to another approach to classical solutions, using the  `Polyakov' action.

\section{Another Approach to Classical Solutions:  `Polyakov' Action}
\label{polyasection}
Although we could quantize a limited subsector of the moduli space of supersymmetric solutions above, the symplectic form and Hamiltonian on the entire moduli space do not lend themselves to simultaneous diagonalization in any simple fashion. So, we will now present another approach to analyzing classical solutions in global AdS that will be useful for quantization.

In the action, \eqref{probebraneeff} that governs the motion of D-string, we can introduce a worldsheet metric to get rid of $\sqrt{-h^{\rm eff}}$. We can then fix conformal gauge and introduce light-cone coordinates $x^{\pm} = \tau \pm \sigma$ to obtain the action
\begin{equation}
\label{Polyakov}
S_{\cal P} = {1 \over 2 \pi} \int (G^{\rm eff}_{\mu \nu} + {B_{\mu \nu} \over \alpha'}) \partial_{+}{X^{\mu}}\partial_{-}{X^{\nu}} \, d x^{+} d x^{-} . 
\end{equation}
This is exactly the same as the usual transition from the Nambu-Goto to the Polyakov action (the ${\cal P}$ stands for Polyakov) for the F-string. We emphasize that the manipulation above is purely classical. 

A classical solution of the action above is equivalent to a classical solution of the DBI action only after we impose the Virasoro constraints:
\begin{equation}
\label{virconstraints}
T(x^+) = \tilde{T}(x^-) = 0 . 
\end{equation}
where $T$ and $\tilde{T}$ are the classical left and right moving stress tensors derived from the action \eqref{Polyakov}. 

The symplectic structure on the the set of all solutions to the action \eqref{Polyakov} that obey the constraints \eqref{virconstraints}, saturate the bound \eqref{bpsbound} and for which the 
worldsheet determinant never vanishes, is identical to the symplectic structure on the set of solutions to the action \eqref{probebraneeff} saturating the bound \eqref{bpsbound}. 
As we explained in the previous section, the symplectic structure on the space of supersymmetric $(p,q)$ strings is the same as the symplectic structure on the space of supersymmetric solutions to the action \eqref{probebraneeff} with the substitutions 
\begin{equation}
\label{substitutions}
Q_5 \rightarrow p(Q_5 - q) + q(Q_1 - p),~~M_{\rm int} \rightarrow {\cal M}_{\rm p,q} .
\end{equation}
This means that, as long as we are interested only in supersymmetric solutions we can use the action \eqref{Polyakov}, with the substitutions \eqref{substitutions}  for $(p,q)$ strings also. This allows us to treat $(1,0)$ strings on the same footing as all other $(p,q)$ strings in the discussion below. 

\subsection{The $SL(2,R) \times SU(2)$ WZW model: Background and Notation}
\label{machinery}
The action \eqref{Polyakov} may be recast as an $SL(2,R) \times SU(2)$ WZW model in addition to the non-linear sigma model on the internal manifold. To see this define,
\begin{equation}
\label{groupparametrization}
\begin{split}
g_1 &= e^{i {t - \theta \over 2} \sigma_2} e^{\rho \sigma_3} e^{i {t + \theta \over 2} \sigma_2} , \\
g_2 &= e^{i {\phi_1 - \phi_2 \over 2} \sigma_3 } e^{i \zeta \sigma_2} e^{i {\phi_1 + \phi_2 \over 2} \sigma_3} .
\end{split}
\end{equation}
Clearly, $g_1 \in SL(2,R)$ and $g_2 \in SU(2)$. The action \eqref{Polyakov}, with the generalization \eqref{substitutions} may be written as:
\begin{equation}
\label{generalpolya}
S = {-k \over 4 \pi} \int {\rm Tr}\{(g_1^{-1} \partial_{\mu} g)^2 + (g_2^{-1} \partial_{\mu} g)^2 \} \, d^2 x + \Gamma_{WZ}^{SU(2)} + \Gamma_{WZ}^{SL(2,R)} + S_{\rm int} ,
\end{equation}
where the terms $\Gamma_{WZ}^{SU(2)}$ and $\Gamma_{WZ}^{SL(2,R)}$ are the usual Wess Zumino terms for $SU(2)$ and $SL(2,R)$ respectively (see \cite{Witten:1983ar} and references therein for details) and $S_{\rm int}$ is the action for the non-linear sigma model on the internal manifold ${\cal M}_{p,q}$. We will, sometimes, find it convenient to work with the group element
\begin{equation}
\label{slsudecomp}
g = g_1 \otimes g_2 ,
\end{equation}
where $g \in SL(2,R) \times SU(2)$.

So, apart from the non-linear sigma model on ${\cal M}_{p,q}$, we now have exactly a WZW model of level $k$ on $SL(2,R) \times SU(2)$. The $SU(2)$ WZW model 
has been studied very widely, and 
the $SL(2,R)$ model has attracted attention in the studies of fundamental strings propagating on $AdS_3$. In what follows, we will draw heavily on the studies of \cite{Giveon:1998ns,Maldacena:2000hw,Maldacena:2000kv}.

It is important that we wish to study the WZW model on the global cover of $SL(2,R)$. In our analysis, we will need to ensure that the string worldsheet closes in global $AdS_3$ and not just in the group parameterization \eqref{groupparametrization}. This has consequences that we will mention below.  

Classical solutions of the WZW model can be decomposed into a product of a left-moving solution and a right-moving solution.
\begin{equation}
\label{classicalwzw}
 g(x^{+},x^{-}) = g^{+}(x^{+}) g^{-}(x^{-}) . 
\end{equation}
The entire solution must, of course, be periodic as a function of $\sigma$, but the two individual components only need to come back to each other up to a {\em monodromy}, $M \in SL(2,R) \times SU(2)$.
\begin{equation}
\label{monodromydef} 
\begin{split}
g^{+}(x^{+} + 2 \pi) &= g^{+}(x^{+}) M ,\\ g^{-}(x^{-} - 2 \pi) &=
M^{-1} g^{-}(x^{-}) .\end{split}
\end{equation} 
The decomposition of equation \eqref{classicalwzw} is not unique. Given a classical solution $g(x^+, x^-)$, a decomposition $\{g^{+}(x^+), g^{-}(x^-)\}$, and any constant group element $U$, one obtains another decomposition of the {\em same} solution via $\{g^{+} U, U^{-1} g^{-}\}$.  Under this $M \rightarrow U^{-1} M U$. Hence, $M$ is determined only up to conjugation. Classical solutions of the WZW model may be classified by the conjugacy class of $M$.
 
The quantum WZW model has a current algebra symmetry and the Hilbert space breaks up into representations of this algebra. It was shown in \cite{Chu:1991pn} that, at least for the case of the $SU(2)$ affine algebra,  all states in a particular representation have the same monodromy eigenvalue. Conversely, as we will do, one may use the monodromy to obtain information about which states occur in the spectrum.

Our model, has six right moving and six left moving conserved currents. Three correspond to $SL(2,R)$ generators, and three correspond to $SU(2)$ generators. Explicitly, these currents are given by
\begin{equation}
\label{currentdefine}
\begin{split}
J^{a}(x^+) &= {k}{\rm Tr}( G^a \partial_{+} g_1 g^{-1}_1) ,~~ \tilde{J}^{a}(x^-) = {k } {\rm Tr}((G^a)^* g^{-1}_1 \partial_{-} g_1) ,\\
K^{i}(x^+) &= {k } {\rm Tr}({-i \sigma^i \over 2} \partial_{+} g_2 g^{-1}_2),~~\tilde{K}^{i} (x^-) = {k } {\rm Tr}({-i (\sigma^i)^* \over 2} g^{-1}_2 \partial_{-} g_2).\\
\end{split}
\end{equation}
In the first line, $a$ runs over the set $\{z, +, -\}$ and we take $G^z = {-i \sigma^y \over 2} , G^{\pm} = G^x \pm i G^y = {1 \over 2} (\sigma^z \pm i \sigma^{x})$. In the second line, $i$ runs over $x,y,z$. 
The left and right moving stress energy tensors are given by
\begin{equation}
\label{stress}
\begin{split}
T(x^+) &= {1 \over k} (-(J^z)^2 + (J^x)^2 + (J^y)^2 + (K^x)^2 + (K^y)^2 + (K^z)^2) + T_{\rm int}(x^+), \\
\tilde{T}(x^-) &= {1 \over k} (-(\tilde{J}^z)^2 + (\tilde{J}^x)^2 + (\tilde{J}^y)^2 + (\tilde{K}^x)^2 + (\tilde{K}^y)^2 + (\tilde{K}^z)^2) + \tilde{T}_{\rm int}(x^-),
\end{split}
\end{equation}
where $T_{\rm int}(x^+), \tilde{T}_{\rm int} (x^-)$ are the left and right moving stress energy tensors of the sigma model on the internal manifold. We will only need the property that $\int T_{\rm int}(x^+) d x^{+} \geq 0, \int \tilde{T}_{\rm int}(x^-) dx^- \geq 0$. 

We will find it convenient to use the modes
\begin{equation}
\label{modes}
\begin{split}
 T(x^+) &= \sum {L_n} e^{i n x^+} ,~~ J^i(x^+) = \sum {J^i_n} e^{i n x^+} ,~~K^i(x^+) = \sum {K^i_n} e^{i n x^+} , \\
\tilde{T}(x^-) &= \sum {\tilde{L}_n} e^{i n x^-} ,~~ \tilde{J}^i(x^-) = \sum {\tilde{J}^i_n} e^{i n x^-} ,~~ \tilde{K}^i(x^-) = \sum {\tilde{K}^i_n} e^{i n x^-} .
\end{split}
\end{equation}

The energy $E$ and angular momentum $L$ in global AdS are related to the zero modes of these currents.
\begin{equation}
\label{relation}
{E + L \over 2} = J^z_0,~~ {J_1 - J_2 \over 2} = K^z_0,~~ {E - L \over 2} = \tilde{J}^z_0,~~ {J_1 + J_2 \over 2} = \tilde{K}^z_0. 
\end{equation}

Hence, the BPS bound \eqref{bpsbound} is saturated when:
\begin{equation}
\label{susy}
\tilde{J}^z_0 = \tilde{K}^z_0 .
\end{equation}

\subsection{Solving the Right-Moving Sector}
\label{polyasolutions}
We will now show that the supersymmetry relation \eqref{susy} and the Virasoro constraints \eqref{virconstraints}, are enough to solve for the entire right-moving sector of the non-linear sigma model \eqref{Polyakov}. 

First, recall that even in conformal gauge, we have the freedom to redefine $x^{-} \rightarrow f(x^{-})$. We will choose this freedom to set 
\begin{equation}
\label{constgauge}
\tilde{J}^z(x^-) = \tilde{J}^z_0, ~(\rm a~constant)
\end{equation}
Let us see how this gauge may be reached. From the definition of the current, \eqref{currentdefine}, we see that under a coordinate transformation $x^-_{\rm old} \rightarrow x^-$:
\begin{equation}
\tilde{J}^z(x^-_{\rm old}) = {\partial x^{-} \over \partial x^-_{\rm old}} \tilde{J}^z(x^-) .
\end{equation}
Hence, if we define a new coordinate by
\begin{equation}
\label{transform}
{\partial x^- \over \partial x^-_{\rm old}} = {\tilde{J}^z(x^-_{\rm old}) \over \tilde{J}^z_0} , 
\end{equation}
we will explicitly reach the gauge \eqref{constgauge}. Notice that \eqref{transform} is always well-defined since to obtain a solution to the Virasoro constraints, we must have $\tilde{J}^{z}_0 > 0$. Second, the constant $\tilde{J}^z_0$ is automatically determined by demanding that the new coordinate have the same periodicity as the old coordinate i.e. $x^{-}(x^-_{\rm old} + 2 \pi) = x^{-}(x^-_{\rm old}) + 2 \pi$. 
This is reassuring, since \eqref{relation} tells us that $\tilde{J}^z_0$ is a physical quantity; so, gauge fixing should leave it unaltered.

Now, consider the Virasoro constraint
\begin{equation}
\tilde{L}_0 = 0 .
\end{equation}
In the gauge above, this reads
\begin{equation}
-(\tilde{J}^{z}_0)^2 + k \tilde{L}_0^{\rm int} + \sum_{n \geq 0} |\tilde{J}^x_{n}|^2 + |\tilde{J}^y_n|^2 + |\tilde{K}^z_{n}|^2 + |\tilde{K}^x_{n}|^2 + |\tilde{K}^y_{n}|^2 = 0 .
\end{equation}
Using relation \eqref{susy}, we find that this implies that, except for $\tilde{J}^z_0$ which is set equal to $\tilde{K^z_0}$ by \eqref{susy} and remains an arbitrary parameter, all the other Fourier components that appear in the expression above are set to zero! 
\begin{equation}
\label{allzero}
\begin{split}
\tilde{L}_0^{\rm int} &= 0 ,\\
\tilde{J}^x_{n} &= \tilde{J}^y_{n} = \tilde{K}^x_{n} = \tilde{K}^y_{n} = 0 ,\\
\tilde{K}^z_{n \neq 0}&=\tilde{J}^z_{n \neq 0}=0 , \\
\tilde{K}^z_0 &=\tilde{J}^z_0 . \\
\end{split}
\end{equation}

We can now solve the equations \eqref{currentdefine} to completely obtain the right moving sector of our theory in terms of the single arbitrary parameter $\tilde{J}^z_0$.  In particular, referring to the notation of \eqref{slsudecomp}, we see that
\begin{equation}
\label{solveright}
\begin{split}
g_1(x^+,x^-) &=   g_1(x^+) \exp\left\{i {\tilde{J}^z_0 \over k} \sigma_2 x^-\right\} ,\\
g_2(x^+, x^-) &= g_2(x^+) \exp\left\{i {\tilde{J}^z_0 \over k} \sigma_3 x^-\right\} .
\end{split}
\end{equation}
All right moving excitations on the internal manifold are also set to zero by \eqref{allzero}. 

Actually, these solutions are just the solutions \eqref{explicitsolutions} in a new guise. Referring to the group parameterization \eqref{groupparametrization}, we see the solutions \eqref{solveright} translate to:
\begin{equation}
\begin{split}
t(\sigma, \tau) &= t(x^+) + {\tilde{J}_0^z \over 2 k} x^-,~~ \theta(\sigma,\tau) = \theta(x^+) + {\tilde{J}_0^z \over 2 k} x^-,~~ \phi_1(\sigma,\tau) =  \phi_1(x^+) + {\tilde{J}_0^z \over 2 k} x^- ,\\
\phi_2(\sigma,\tau) &=  \phi_2(x^+) + {\tilde{J}_0^z \over 2 k} x^-,~~\rho(\sigma,\tau) = \rho(x^+),~~ \zeta(\sigma,\tau) = \zeta(x^+),~~ z^a(\sigma,\tau) = z^a(x^+) .
\end{split}
\end{equation}

Two points are worth emphasizing. 
\begin{enumerate}
\item
By solving the right-moving side of the $SL(2,R)$ and  $SU(2)$ WZW models, we have also determined the monodromy of the left-moving side. 
\item
The monodromy of the $SU(2)$ part and the $SL(2,R)$ part are linked, since they both depend on the same parameter $\tilde{J}^z_0$.  
\end{enumerate} 

As we mentioned, the monodromy of the solution gives us information about which representation of the current algebra we are in.  The two features above then mean that at least semi-classically, once we specify the representation of the right-moving $SL(2,R)$ current algebra this determines the representation of the left-moving $SL(2,R)$ algebra and the left and right moving $SU(2)$ current algebras(the inclusion of fermions modifies this statement slightly as we discuss in Section \ref{spacetimepartfun}).

$SL(2,R)$ has three types of conjugacy classes. Given $\Gamma \in SL(2,R)$, we determine its conjugacy class to be of type elliptic ($({\rm tr}(\Gamma))^2 < 4$), parabolic ($({\rm tr}(\Gamma))^2=4$) or hyperbolic ($(tr(\Gamma))^2 > 4$). We see that, generically, the monodromy of the solutions \eqref{solveright} lies in an {\em elliptic} conjugacy class of the group. We will find, later, that quantizing these solutions will give rise to `short strings' in $AdS_3$. This is linked to the observation made above that unless $E-L$ is quantized in units of $k$, our strings are bound to the center of $AdS_3$.
When $\tilde{J}^{z}_0 = {n k \over 2}$ in equation \eqref{solveright} for some integer $n$, the monodromy of the solutions \eqref{solveright} is ${\pm 1}$. This
kind of solution can escape to infinity and lies at the cusp of short and long strings. One may suspect that on quantization these solutions would give rise to states at the bottom of a continuum. Semi-classically, this is indeed true. The full quantum analysis in Section \ref{spacetimepartfun} raises a puzzle regarding this that we will discuss there.

\subsection{Winding Sectors}
Notice, that given a solution of the form \eqref{solveright}  we can generate another solution using the transformation
\begin{equation}
\begin{split}
\label{classpectflow}
g_1(x^+, x^-) &\rightarrow e^{i w_1 \sigma_2 {x^+ \over 2}} g_1 e^{i w_1 \sigma_2 {x^{-} \over 2}} ,\\
g_2(x^+, x^-) &\rightarrow e^{i w_2 \sigma_3 {x^+ \over 2}} g_2 e^{i w_1 \sigma_3{x^{-} \over 2}} .
\end{split}
\end{equation}
In equation \eqref{solveright}, this operation takes $\tilde{J}^z_0 \rightarrow \tilde{J}^z_0 + {k w \over 2}$. 

The two parameters that determine the `spectral flow' operation above have the property that $w_1, w_2 \in \mathbb{Z}$ and $w_2 = w_1 {\rm (mod)} 2$. Notice, two important features above. First, we have to spectral flow the left moving part of the $SL(2,R)$ model by exactly the same amount as the right moving part; this is required by the periodicity of the worldsheet in global $AdS_3$ which is the global cover of $SL(2,R)$. Supersymmetry now determines that the right-moving part of the $SU(2)$ WZW model must also be spectrally flowed by $w_1$. However, periodicity on $S^3$ merely requires $w_2 = w_1 {\rm (mod)} 2.$

Second, since $\pi_1(SU(2)) = 0$, we cannot classify the solutions of the $SU(2)$ WZW model by their winding number. This is not true for $SL(2,R)$ since $\pi_1(SL(2,R)) = \mathbb{Z}$. Solutions to the $SL(2,R)$ WZW model hence break up into sectors labelled by two integers, one for the left-moving solution and the other for the right-moving solution. Since we are considering the global cover of $SL(2,R)$, closure of the worldsheet requires the two integers to be equal.  So,solutions of the WZW model with target space the global cover of $SL(2,R)$ break up into sectors labelled by an single integer $w$.

\subsection{${1 \over 2}$ BPS states}
\label{chiralchiralsection}
Before we conclude our discussion of classical solutions, we would like to discuss two additional issues. The first regards `chiral, chiral primaries' in global AdS. These are half-BPS states of the $N=4$ algebra on the boundary and are chiral primaries on the left and on the right. This means that they satisfy the BPS relations
\begin{equation}
\label{chiralchiralbps}
\begin{split}
E - L &= J_1 + J_2 ,\\ 
E + L &= J_1 - J_2 .
\end{split}
\end{equation}

Extending the analysis of section \ref{review}, we conclude that probes that maintain, both 
\begin{equation}
\begin{split}
n_1 &= {\partial \over \partial t} + {\partial \over \partial \theta} + {\partial \over \partial \phi_1} + {\partial \over \partial \phi_2} ,\\
n_2 &= {\partial \over \partial t} - {\partial \over \partial \theta} + {\partial \over \partial \phi_1} - {\partial \over \partial \phi_2} , 
\end{split}
\end{equation}
preserve the required 8 supersymmetries. 

From our list of solutions, we can see that this fixes both the $\sigma$ dependence and the $\tau$ dependence. In particular, the only allowed solutions are:
\begin{equation}
\label{giantchiralchiral}
\begin{split}
t &= \tau ,~~~ \theta = w \sigma + \tau ,~~~ \phi_1 = {\rm const} + \tau ,~ \phi_2 = w \sigma + \tau , \\
\rho &= {\rm const} ,~~~ \zeta = {\rm const},~~~z^i = {\rm const} .
\end{split}
\end{equation}
The two tangent vectors above are then, ${\partial \over \partial \tau}$ and ${\partial \over \partial \tau} - {2 \over w} {\partial \over \partial \sigma}$. 

We now encounter a surprise. Calculating the charges of these solutions from table \ref{global}, we find that for a $(p,q)$ probe, $E = J_1 = {k w}$ and $L = J_2 = 0$ where $k$ is given by \eqref{kdef}. However, the boundary theory has chiral-chiral primaries for all half-integer values of the scaling dimension(${E+L \over 2}$) upto ${Q_1 Q_5 \over 2}$, not just the small subset above. Evidently, smooth giant gravitons cannot describe generic chiral primaries; a point that was stressed in \cite{Lunin:2002bj}. 

Moving now to the `Polyakov' approach, we can find the classical solutions for chiral-chiral primaries by merely repeating the analysis above for the left-moving side.  We find that the solutions that obey the relations \eqref{chiralchiralbps} and have the correct periodicity on the worldsheet are:
\begin{equation}
\label{chiralchiral}
\begin{split}
g_1(x^+,x^-) &= \exp\left\{i {J^z_0 \over k} \sigma_2 x^+\right\}  \exp\left\{i {J^z_0 \over k} \sigma_2 x^-\right\} ,\\
g_2(x^+, x^-) &= \exp\left\{i {J^z_0 \over k} \sigma_3 x^+\right\} \exp\left\{i {J^z_0 \over k} \sigma_3 x^-\right\} , 
\end{split}
\end{equation}
where, as before, $g_1$ is an element of $SL(2,R)$ and $g_2$ an element of $SU(2)$. The form \eqref{chiralchiral} is unique up to an irrelevant additive shift in $\sigma$. If we move to spacetime, using the parameterization \eqref{groupparametrization}, we find that the solutions \eqref{chiralchiral} correspond to curves that pass through $\rho = 0$ (the center of $AdS$) and sit at $\zeta = 0$.\begin{footnote}{The Polyakov formalism can accommodate these solutions because solutions to the action \eqref{Polyakov} obeying the constraints \eqref{virconstraints} comprise all solutions to the action \eqref{probebraneeff} plus geodesics}\end{footnote} Hence, the values of $\theta$ and $\phi_2$ are ill defined but if we take these values to be zero, then the solutions \eqref{chiralchiral} correspond to $t = {2 J^z_0 \tau \over k},~\theta = 0,~ \rho = 0,~\phi_1 = {2 J^z_0 \tau \over k},~\phi_2 = 0,~\zeta=0$. 

We now notice a remarkable feature about these solutions. Spectral flow does not puff these geodesics into strings! The transformation \eqref{classpectflow} takes $J^z_0 \rightarrow J^z_0 + {w k \over 2}$ but leaves the solution in the form of a geodesic placed at $\rho = 0, \zeta = 0$.
 This simple observation explains several facts about the bulk spectrum of chiral-chiral primaries that have hitherto been puzzles:
\begin{enumerate}
\item
The spectrum of chiral-chiral operators in non-zero winding sectors was calculated in a nice paper by Argurio, Giveon and Shomer \cite{Argurio:2000tb} (AGS) and found to be a continuation of a graviton spectrum. While one may expect stringy effects to start showing at energies of order $Q_5$ or $Q_1$,  this does not happen for chiral-chiral operators because spectral flow does not puff these geodesics up into strings. This also explains why de Boer, in \cite{deBoer:1998us}, was successful in reproducing the spectrum of (1/2) BPS states on the boundary up to energies ${Q_1 Q_5 \over 2}$ by naively extending the graviton spectrum.
\item
AGS conjectured that in each winding sector, some chiral-operators (with charges integrally quantized in units of ${Q_5 \over 2}$) vanished into the continuum. We see, that at exactly these values of the charge, chiral-chiral primaries are described by giant gravitons as in equation \eqref{giantchiralchiral}. Classically, they can be at any value of $\rho$ including $\rho \rightarrow \infty$. Quantum mechanically this means that they are at the bottom of a continuum of non-supersymmetric states and we may expect difficulty in counting them.
\item
From the spectrum of chiral operators, AGS also discussed the possibility that the boundary theory was a deformation of the iterated symmetric product  $((M_{int})^{Q_5}/S_{Q_5})^{Q_1}/S_{Q_1}$. However, since the classical solutions corresponding to chiral operators are geodesics they do not differentiate between different probes; we cannot determine if they are constituted by D1 branes, D5 branes or a bound state of these. Even in the semi-classical quantum analysis below we find that chiral-chiral operators can be obtained by quantizing any of these probes.This restores the democracy between $Q_1$ and $Q_5$.
\item
Correlation functions of chiral-chiral operators, in the zero-winding sector, were recently calculated by Gaberdiel and Kirsch \cite{Gaberdiel:2007vu} and Dabholkar and Pakman \cite{Dabholkar:2007ey}. The insight above, that chiral-chiral operators do not `see' winding, indicates that similar results would be obtained by repeating this calculation in sectors of non-zero winding.
\end{enumerate}

\section{Semi-Classical Quantization}
\label{semiclassicalsection}
We will now use the insights of the previous sections to deduce features of the ${1\over 4}$ BPS sector of quantum string theory on $AdS_3$. Throughout this section, we will work in a semi-classical limit, where the charges of the states that we consider are large and hence, for example $j(j+1)$ may be well approximated by $j^2$. There are two reasons for doing this. The first is that the analysis we perform here is valid for general $(p,q)$ probes. The second is we will find that when we perform an exact analysis of the D-string by dualizing to a F1-NS5 frame, it will turn out the formulae we derive in this section are {\em quantitatively} correct including all the additive factors of 1. The factors that we neglect, conspire to cancel!

As we mentioned in the previous section, a $(p,q)$ probe leads to an $SL(2,R) \times SU(2)$ model with level $k$ given by equation \eqref{kdef}. The details of the internal manifold, ${\cal M}_{p,q}$ will not be too important for us here. 

The approach we will adopt is as follows. We start by reviewing the Hilbert space of the $SL(2,R)$ and $SU(2)$ WZW models. 
The question that faces us, then is to understand what sector of this Hilbert space corresponds to the solutions \eqref{solveright} obeying the Virasoro constraints. We tackle this question in subsection \ref{linkingclassicalquantum}.

\subsection{The SU(2) and SL(2,R) WZW models: a review}
The $SL(2,R)$ and $SU(2)$ WZW models each have 3 left moving conserved currents defined in equation \eqref{currentdefine} which we call $J^i$ and $K^i$ respectively. In the quantum theory these currents give rise to an affine symmetry via the commutation relations:
\begin{equation}
\label{explicitsltwor} \begin{split} 
[J^{z}_{n},J^{\pm}_{m}] &= {\pm} J^{\pm}_{n+m},~~  [J^{z}_{n}, J^{z}_m] = -{k n \over 2} \delta_{n+m,0},~~ [J^{+}_{n}, J^{-}_m] =  -2 J^{z}_{n + m} + k n \delta_{n + m, 0}.  \\
 [K^{z}_{n}, K^{\pm}_{m}] &= \pm K^{\pm}_{n+m},~~ [K^{z}_{n}, K^{z}_m] = {k n \over 2} \delta_{n+m,0},~~ [K^{+}_{n}, K^{-}_m] =  2 K^{z}_{n + m} + k n \delta_{n + m, 0}. 
\end{split}
\end{equation}
The Stress Energy tensors, for each algebra, are given by the usual Sugawara construction. In particular, the modes $L_n$ are given by:
\begin{equation}
\begin{split}
L_n^{SL(2,R)} &= {1 \over 2(k-2)} {\bf :} \left[ \sum_{m=-\infty}^{+\infty} J^{+}_{m} J^{-}_{n-m} + J^{-}_{m}J^{+}_{n-m} -  2 J^{z}_{m}J^{z}_{-m} \right] {\bf :} \\
L_n^{SU(2)} &= {1 \over 2(k+2)} {\bf :} \left[\sum_{m=-\infty}^{+\infty}K^{+}_{m} K^{-}_{n-m} + K^{-}_{m} K^{+}_{n-m} + 2 K^{z}_{m}K^{z}_{-m} \right]:
\end{split}
\end{equation}
where ${\bf :}~ \ldots ~{\bf :}$ implies normal ordering where negatively moded operators moded are placed before positively moded operators. These modes obey the algebra:
\begin{equation}
\label{sugawaralgebra}
\begin{split}
[L_n^{SL(2,R)}, J_m^a] &= -m J^{a}_{n+m}, ~[L_n^{SU(2)}, K_m^a] = -m K^{a}_{n+m}, \\
[L_n^{SL(2,R)}, L_m^{SL(2,R)}] &= (n-m)L^{SL(2,R)}_{n+m} + {k \over 4(k-2)} (n^3 - n) \delta_{n+m,0}\\
[L_n^{SU(2)},L_m^{SU(2)}] &= (n-m)L^{SU(2)}_{n+m} + {k \over 4(k+2)} (n^3 - n) \delta_{n+m,0}
\end{split}
\end{equation}

Representations of the $SU(2)$ affine algebra are constructed by starting with a lowest weight state $|j  \rangle$ obeying $K^{\pm, z}_{n>0} |j  \rangle=0,~~ K^{+}_{0} |j  \rangle = 0,~~ K^{z} |j  \rangle = j |j  \rangle $. Such a state is called an `affine primary'. 
Given an affine primary, one acts in all possible ways with the lowering operators $K^{\pm,z}_{n<0}, K^{-}_{0}$ and removes null states to construct the entire representation.  We will denote a representation built on an affine primary of weight $j$ as ${\cal L}^j$ 
The spectrum of the $SU(2)$ model at level $k$ comprises the `diagonal modular invariant' $\oplus_{j = 0, {1 \over 2} \ldots {k \over 2}} {\cal L}^j \otimes \bar{\cal L}^j$, where the left moving affine primary has the same weight as the right moving affine primary. We refer the reader to \cite{DiFrancesco:1997nk} for details.

The spectrum of the $SL(2,R)$ WZW model is more intricate, because this group is non-compact. The $SL(2,R)$ WZW model also  has lowest weight representations of the kind described above. These are discussed in \cite{Dixon:1989cg,Maldacena:2000hw} and we refer the interested reader there for details. Here, we review the two kinds of representations that are most relevant to strings propagating on $AdS_3$. 
\begin{enumerate}
\item
Discrete Lowest Weight Representations ${\hat{{\cal D}}}_{j}^+$ : These representations are labeled by a real number $j$. $j$ is related to the second Casimir via $c_2 = {1 \over 2} \{J_0^{+}, J_{0}^{-}\} - (J_0^z)^2 = -j(j-1)$. One starts with a state $|j,j \rangle$ obeying
\begin{equation}
J^{\pm, z}_{n>0} |j,j \rangle =0 ,~~ J^{-}_{0} |j,j \rangle = 0 ,~~ J^{z}_{0} |j, j \rangle = j|j,j \rangle , 
\end{equation}
and then acts with the remaining operators of the algebra $\{J^{\pm, z}_{n < 0}, J^{+}_{0}\}$ to obtain the entire representation. 
\item
Continuous Lowest Weight Representations ${\hat{\cal C}}_j^{\alpha}$ : These representations are labelled by a real number $s$ with $j = {1 \over 2} + i s$. The second Casimir $c_2 = -j(j-1) = s^2 + {1 \over 4}$. One starts with a state $|s, \alpha, \alpha \rangle$ obeying
\begin{equation}
J^{\pm, z}_{n>0} |j,\alpha, \alpha \rangle =0 ,~~ J^{z}_{0} |j, \alpha, \alpha \rangle = \alpha|j,\alpha,\alpha \rangle ,
\end{equation}
and acts with the remaining operators of the algebra  $\{J^{\pm, z}_{n < 0}, J^{\pm}_{0}\}$ to obtain the entire representation. Evidently, one may restrict $0 \leq \alpha<1$
\end{enumerate}

Now, notice that both the $SL(2,R)$ and $SU(2)$ models have a `spectral flow' symmetry. The transformations  
\begin{equation}
\label{spectralflow}
\begin{split}
J^{z}_{n} &\rightarrow J^{z}_{n} + {k w \over 2} \delta_{n, 0} ,~~ K^{z}_{n} \rightarrow K^{z}_{n} + {k w \over 2} \delta_{n, 0} \\
J^{\pm}_n &\rightarrow J^{\pm}_{n \mp w} ,~~ K^{\pm}_n \rightarrow K^{\pm}_{n \pm w}, \\
L^{SL(2,R)}_n &\rightarrow L^{SL(2,R)}_{n} - w J^{z}_n - {k w ^2 \over 4} \delta_{n,0},~~L^{SU(2)}_n \rightarrow L^{SU(2)}_{n} + w K^{z}_n + {k w ^2 \over 4} \delta_{n,0}
\end{split}
\end{equation}
 preserve the algebra \eqref{explicitsltwor} and \eqref{sugawaralgebra}.\begin{footnote}{In this paper, we will think of spectral flow as an operation on states that leaves the operators themselves unchanged}\end{footnote} For the $SU(2)$ algebra, at level $k$, spectral flow by an odd number of units maps us from a representation of lowest weight $j$ to a representation of lowest weight ${k \over 2} - j$. Spectral flow by an even number of units maps us back to the representation of lowest weight $j$. 
However, for the $SL(2,R)$ algebra, spectral flow generically produces a new representation that is not a lowest weight representation at all! We denote these spectrally flowed representations by ${\hat{\cal{D}}}^{w,+}_j$ and $\hat{\cal C}^{w,\alpha}_{j}$. It was explained first in \cite{Henningson:1991jc} and later in \cite{Maldacena:2000hw,Maldacena:2000kv} that a consistent Hilbert space of bosonic strings propagating in $AdS_3$ is formed by considering all $\hat{\cal{D}}^{w,+}_j \otimes {\hat{\bar{\cal D}}}^{w,+}_j$ with ${1 \over 2} < j < {k - 1 \over 2}$ and all ${\hat{\cal C}}^{w,\alpha}_{{1 \over 2} + i s} \otimes {\hat{\bar{\cal C}}}^{w,\alpha}_{{1 \over 2} + i s}$. Note that the value of $j$ and $w$ on the right and left have to be the same.

\subsection{Linking Classical Solutions to Quantum States}
\label{linkingclassicalquantum}

We need to identify which subsector of the spectrum above corresponds to the solutions discussed in Section \ref{polyasolutions}. Recall that an analysis of supersymmetry allowed us to completely solve the right-moving sector. This, in turn, also determined the monodromy of the left-moving sector, since the left and right moving parts of the solution are constrained to have the same monodromy. The link between classical solutions and quantum states goes through the monodromy. The key result that we need was proved by Chu et. al. in  \cite{Chu:1991pn} drawing on the study of \cite{Felder:1988as}. For other studies examining the canonical formalism applied to WZW models, see \cite{Gawedzki:1990jc} and references therein.

Recall that the phase space of the WZW model consists of all classical solutions to the action and these are of the form \eqref{classicalwzw}. The conjugacy class of the monodromy is a well defined function on phase space. Canonical quantization promotes this function to an operator. The authors of \cite{Chu:1991pn} considered the $SU(2)$ model. Conjugacy classes of $SU(2)$ are labelled by a single real number $0 \leq \nu <  \pi$ with corresponding group element $e^{i \nu \sigma_3}$. In \cite{Chu:1991pn}, it was shown that states in the representation ${\cal L}^j$ with $0 < j < {k \over 2}$ were eigenstates of the operator $\nu$ with eigenvalue
\begin{equation}
\label{monodromyprimary}
\nu= {2 j + 1 \over k+2} \pi. ~~ {\rm [SU(2)]}
\end{equation}

The analysis of \cite{Chu:1991pn} is rather intricate but it is not hard to understand the semi-classical origins of formula \eqref{monodromyprimary}. The affine primary of a lowest weight representation, and other states obtained by acting on it with the zero-modes of the algebra, are the states in the representation that have the lowest conformal weight. Hence, we can derive a semi-classical version of formula \eqref{monodromyprimary} by considering all solutions with a given monodromy and minimizing their conformal weight. 
Consider, the right moving part of a classical solution of the $SU(2)$ WZW model, which we parameterize as in \eqref{groupparametrization}
\begin{equation}
g_2(x^-) =  e^{-i {\phi_1(x^-) - \phi_2(x^-)\over 2} \sigma_3} e^{i \zeta(x^-) \sigma_2} e^{-i {\phi_1(x^-) + \phi_2(x^-) \over 2} \sigma_3},
\end{equation}
with the boundary condition:
\begin{equation}
\label{bdcondition}
g_2(x^- + 2 \pi) = e^{i \sigma_3 \nu} g_2(x^-).
\end{equation}
We can obtain the currents of this group element using the formulae in \eqref{currentdefine}. We find that the zero-modes of the stress energy tensor and $K^z$ current are given by
\begin{equation} 
\label{conformalweight}
\begin{split}
L^{\rm SU(2)}_0 &= {1 \over 2 \pi k} \int_0^{2 \pi} \left(\cos^2{\zeta} (\phi_1')^2 + \sin^2{\zeta} (\phi_2')^2 + (\zeta')^2 \right) d x^- ,\\
K^z_0&= {k \over 2 \pi} \int \left( \cos^2{\zeta} \phi_1' + \sin^2{\zeta} \phi_2' \right) d x^- .
\end{split}
\end{equation}
If we minimize the conformal weight in \eqref{conformalweight} subject to the boundary condition \eqref{bdcondition} and we find that the minimum is reached at:
\begin{equation}
\label{minenergysolution}
\begin{split}
\zeta &= {\rm constant} ,\\
\phi_1'&= -\phi_2' = {\nu \over 2 \pi} x^{-} ,\\
L_0^{\rm SU(2)} &= {\nu^2 \over (4 \pi)^2 k} .
\end{split}
\end{equation}
If we now use the fact that an affine primary of weight $j$ has conformal weight, ${j(j+1) \over k + 2}$, we find the semi-classical relation
\begin{equation}
\label{semiclassmonosutwo}
\nu \sim {2 j \pi \over k} .
\end{equation}
where the $\sim$ indicates that this relation is semi-classical. Quantum fluctuations will modify this relation to the exact equation \eqref{monodromyprimary}. The formula \eqref{minenergysolution} is valid as long as $\nu \leq \pi$ (otherwise, it is shorter to go around the sphere the `other way'), which is consistent with the fact that the lowest weight affine primaries of the $SU(2)$ affine algebra are capped at $j = {k \over 2}$. 
  
To see the significance of the constant value of $\zeta$ in \eqref{minenergysolution}, we calculate on this solution:
\begin{equation}
K^z_0 = \cos (2 \zeta) {k \nu \over 2 \pi} = \cos(2 \zeta) j.
\end{equation}
The possible values of $K^z_0$ for the lowest conformal weight of \eqref{minenergysolution} range from $[-j, +j]$. This is what we expect since all states in the $SU(2)$ representation built by acting with the {\em zero-modes} of the affine algebra on the affine primary have the same conformal weight. They are distinguished by their eigenvalues under $K^z_0$ and these eigenvalues can range from $-j \ldots j$ for an affine primary of weight $j$. The highest value of $j$ is what corresponds to the affine primary, as we defined it above, and this is obtained at $\zeta=0$. 

Now, we notice a remarkable fact. At this value of $\zeta$, the solution \eqref{minenergysolution} is exactly the form that the right-moving sector, in the zero winding sector, takes in \eqref{solveright}. This suggests that the solutions \eqref{solveright}, for $\tilde{J}^z_0 < {k \over 2}$ correspond to states that, on the right moving side, are affine primaries of $SU(2)$. The solutions with $\tilde{J}^z_0 > {k \over 2}$ can always be obtained from these solutions by means of the spectral flow operation \eqref{classpectflow}. Hence, the solutions with $\tilde{J}^z_0 > {k \over 2}$ correspond to states that are obtained by spectrally flowing an $SU(2)$ affine primary using \eqref{spectralflow}.

We can repeat the semi-classical analysis above for the $SL(2,R)$ affine algebra. Consider a curve in $SL(2,R)$ parameterized by:
\begin{equation}
\label{sltworightparam}
g_1(x^-) = e^{i {t(x^-) +\theta(x^-) \over 2} \sigma_2} e^{\rho(x^-) \sigma_3} e^{i {t(x^-) - \theta(x^-)\over 2} \sigma_2} .
\end{equation}
As we explained above, $SL(2,R)$ has three types of conjugacy classes. However, the solutions of \eqref{solveright} have a monodromy that belongs to an elliptic conjugacy class. Hence, we will consider the boundary condition:
\begin{equation}
\label{bdconditionads}
g_1(x^- + 2 \pi) = e^{i \sigma_2 \nu} g_1(x^-) .
\end{equation}
This time, we have:
\begin{equation}
\begin{split}
L_0 &= {1 \over 2 \pi k } \int_0^{2 \pi} (-\cosh^2{\rho} t'^2 + \sinh^2{\rho} \theta'^2 + \rho'^2) \, d x^- ,\\
J^z_0 &= {k \over 2 \pi } \int_0^{2 \pi} (\cosh^2{\rho} t' - \sinh^2{\rho} \theta'^2) \, d x^-  .
\end{split}
\end{equation}
The group parameterization \eqref{sltworightparam} admits curves that wind around the `time' direction but restricting to the zero winding sector, we would find that the solution that minimizes the conformal weight is:
\begin{equation}
\label{minsltwosolution}
\begin{split}
\rho &= {\rm constant} ,\\
t' &= -\theta' = {\nu \over 2 \pi} x^- .
\end{split}
\end{equation}
This solution has conformal weight, and $J^z_0$ eigenvalue:
\begin{equation}
\label{sltwominweights}
\begin{split}
L_0^{\rm SL(2,R)}=-{\nu^2 \over 4 \pi^2 k},~~J^z_0 = \cosh{2 \rho} {k \nu \over 2 \pi}.
\end{split}
\end{equation}
The value of $L_0^{\rm SL(2,R)}$ above corresponds to the lowest conformal weight possible in a {\em discrete} unflowed representation $\hat{\cal D}^{0,j}$ with
\begin{equation}
\nu \sim {2 j \pi \over k} .
\end{equation}
Repeating the argument of \cite{Chu:1991pn} for the $SL(2,R)$ affine algebra yields the quantum result:
\begin{equation}
\nu = {2 j - 1 \over k - 2} \pi. ~~ {\rm [SL(2,R)]}
\end{equation}
As above, the formula \eqref{minsltwosolution} is valid for $j \leq {k \over 2}$ which tells us that we should consider the discrete unflowed representations $\hat{\cal D}^{0,j}$ only for $j < {k \over 2}$. The value of $J^z_0$ in \eqref{sltwominweights}, can range from $j \ldots \infty$ which is exactly what we expect. The affine primary itself, corresponds to the lowest possible value of $j$ which corresponds to $\rho=0$ in \eqref{minsltwosolution}. At this value of $\rho$, the solution \eqref{minsltwosolution} becomes identical to the right-moving solution of \eqref{solveright}.  Hence, we conjecture that the solutions \eqref{solveright} correspond, on the right-moving side, to states affine primaries of discrete representation of the $SL(2,R)$ affine algebra or to spectral flows of these.

As we mentioned above, there are two other types of conjugacy classes of $SL(2,R)$. The hyperbolic conjugacy classes, in particular, correspond to solutions that have monodromy  $e^{s \sigma_3}$. The minimum conformal weight for classical solutions with this boundary condition is $L_0^{\rm SL(2,R)} = {s^2 \over 4 \pi^2 k}$. Hence, solutions with this monodromy correspond to states in continuous representations. The solutions of \eqref{solveright}, when $J^z_0 = {k \over 2}$ are then at the bottom of a continuum i.e we can reach the continuum by moving infitesimally away from supersymmetry. We will have more to say on this below.

We are now in a position to identify the solutions of \eqref{solveright} with ${1 \over 4}$ BPS states in spacetime. Recall that we are looking for states of the form $|{\rm anything} \rangle|{\rm chiral~primary} \rangle$. The classical solutions of \eqref{solveright} also have this form. They have a very special structure on the right-moving side and an arbitrary solution on the left moving side. It is then natural to conjecture the following
\begin{enumerate}
\item
Left (Right) movers on the worldsheet give rise to left (right) movers in spacetime.
\item
A chiral primary in spacetime is constructed either (a) by taking the affine primary, of a discrete $SL(2,R)$ representation $\hat{{\cal D}}^j$ and combining it with an affine primary of the $SU(2)$ representation ${\cal L}^j$ \begin{footnote}{In the exact analysis of Section \ref{spacetimepartfun}, we find that this construction must be modified slightly. We need to combine the affine primary of $\hat{{\cal D}}^{j+1}$ with the affine primary of  ${\cal L}^j$ and dress the state with fermion zero modes}\end{footnote} or (b) by spectrally flowing a state of this form by $w$ units.
\item
The arbitrary left-moving side of \eqref{solveright} is subject to global constraints from the right-moving side. Semi-classically, we see that this left-moving state must belong to the sector  ${\hat{\cal D}^{j,w}} \times {\cal L}^{j'} \times H^0({\cal M}_{p,q})$ of the sigma model on $SL(2,R) \times SU(2) \times {\cal M}_{p,q}$. Here, $j' = j$, if $w$ is even and $j' = {k \over 2} - j$ otherwise. Of course, we need to impose the left-moving physical state conditions as well.
\item
It appears that at special values of the charges, when the right-moving chiral primary has $j = {k w \over 2}$ (recall that according to point (2) above all such right-moving chiral primaries in spacetime are related by spectral flow on the worldsheet) we obtain states that are the bottom of continua. 
\end{enumerate}

\subsection{Semi-Classical Analysis of Supersymmetric States}
\label{semiclassicalanalysissusy}
We will now verify the conjecture above by checking that the states described above do indeed obey the physical state conditions and BPS relation and also discuss, in more detail, the structure of the arbitrary states that appear on the left-moving side.
Classically, we need to impose the constraints \eqref{virconstraints}. Quantum mechanically, we will demand that physical states $|a  \rangle$ satisfy $L_n |a  \rangle = 0$ and we will mod out by spurious states of the form $|a  \rangle = L_{-n} |b  \rangle$. What about the mass-shell condition? In passing from classical solutions to quantum states, since we are interested {\em only} in the spectrum, we have the freedom to choose normal ordering constants. This issue is discussed in some more detail in the next section. In this subsection, since we are in a regime where all charges are large compared to $1$,  we will not be too precise about this and work with a semi-classical mass-shell condition $L_0|a  \rangle \sim 0$.

First, consider the construction of chiral-primaries in the zero winding sector. Consider a right moving state $|c^{j,0} \rangle$ that is an affine primary of ${\cal L}^j$ and an affine primary of ${\hat{\cal D}}^{0,j}$ and in the ground state in ${\cal M}_{p,q}$. Then,
\begin{equation}
\label{verify}
\begin{split}
 &\tilde{L}_0|c^j \rangle = (\tilde{L}_0^{\rm SL(2,R)} + \tilde{L}_0^{\rm SU(2)}) |c^{j,0} \rangle \sim (-{j^2 \over k}+{j^2 \over k}) |c^{j,0} \rangle = 0 , \\
&\tilde{L}_{n}|c  \rangle = (\tilde{L}_n^{\rm SL(2,R)} + \tilde{L}_n^{\rm SU(2)}) |c^{j,0} \rangle = 0 ,\\
&(\tilde{J}_0^z - \tilde{K}_0^z)|c^{j,0} \rangle = (j - j)|c^{j,0} \rangle = 0 .
\end{split}
\end{equation}
 So, $|c  \rangle$ obeys the physical state and supersymmetry conditions. This state cannot be written as a conformal descendant on the worldsheet. Hence it is not spurious. So, it gives us a good description of a spacetime chiral primary. Semi-classically, it appears that $j$ runs over all the values $0 \leq j < {k \over 2}$ in half-integral steps. The exact analysis of the next section shows us that we actually obtain a series where $j$ runs over ${1 \over 2}, 1, \ldots {k \over 2} - {1 \over 2}$. 

Now, notice that, given a state that satisfies the physical state and supersymmetry conditions above, the transformations \eqref{spectralflow} take us to another state that also satisfies these conditions. So the state $|c^{j,w} \rangle$ obtained by {\em simultaneously} spectral flowing $|c^{j,0} \rangle$ by $w$ units in both $SL(2,R)$ and $SU(2)$ is also a good spacetime chiral primary. 
Note that this process of spectral flow merely extends the series above, from ${1 \over 2} \ldots {k \over 2} - {1 \over 2}$  to ${k w \over 2} + {1 \over 2} \ldots {k (w + 1) \over 2} - {1 \over 2}$ .

This leaves behind gaps at the values ${k w \over 2}$. This is exactly the value of $J^z_0$, where the monodromy of the solutions \eqref{solveright} becomes $1$. It is also the charge assignment for which we explained, in section \ref{review}, that classical probe brane solutions could escape to infinity. It is tempting to believe then, that at these values of $J^z_0$ then, the chiral primaries lie at the bottom of a continuum. We will discuss this further in a moment.

Continuing with our discussion of discrete states, let us denote the state on the left-moving side as $|a \rangle$ (for `arbitrary'). The global constraints of the spectrum described above, mean that that 
\begin{equation}
\label{rightglobal}
|a \rangle \in {\hat{\cal D}}^{j,w} \times {{\cal L}}^{\bar{j}(w)} \times H^{0}({\cal M}_{p,q}) ,
\end{equation}
where $\bar{j}(w) = j$ if $w$ is even and ${k \over 2} - j$ if w is odd.
 In addition, we must impose the physical state conditions
\begin{equation}
\label{physicalarbitrary}
\begin{split}
{L}_n |a  \rangle &=0 ,\\
{L}_0 |a  \rangle &\sim 0 .
\end{split}
\end{equation}
To write an energy formula for $|a  \rangle$, it is convenient to consider the state $|a^{-w} \rangle$ obtained by spectral flowing $|a \rangle$ by $-w$ units. This takes us to the zero-winding sector in $SL(2,R)$ and to the representation ${\cal L}^j$ in $SU(2)$. Note, that the physical state condition implies $L_{n>0} |a^{-w}  \rangle = 0$.  Now, $|a^{-w} \rangle$ may be indexed by its level in $SL(2,R)$, $N$,\begin{footnote}{By `level' here, we mean the oscillator level which is the difference in the conformal weight of the state and the conformal weight of the zero mode. For example, a state $|\Omega \rangle \in  {\hat{\cal D}}^{j,w}$ has level: $(L_0^{\rm SL(2,R)} + {j(j-1) \over k - 2})|\Omega \rangle \equiv N |\Omega \rangle$}\end{footnote}  its level in $SU(2)$,$h_2$, its level in the internal CFT on ${\cal M}_{p,q}$, $h_{int}$ and its $J^z_0$ eigenvalue, $j+Q$ and its $K^z_0$ eigenvalue $j + P$. $Q$ can be negative because, for example, we can act with $J^{-}_{-1}$ on the lowest weight state, but we have the constraint that $Q \geq -N$. $P$ can be negative too, because we can act with $K^{-}_0$ on the lowest weight state.

So, the mass shell condition for $|a \rangle$ then reads:
\begin{equation}
\label{physicalstate}
\begin{split}
&L_0|a  \rangle = \left({-(j+{k w \over 2})^2 \over k} - w Q + {(j + {k w \over 2})^2 \over k} + w P + N + h_2 + h_{\rm int}\right) | a  \rangle = 0 \\
&\Rightarrow Q = P+ {N + h_2 + h_{\rm int} \over w} . 
\end{split}
\end{equation}
Finally, we may write the spacetime charges of the state $|a \rangle |c^{w,j} \rangle$ as
\begin{equation}
\label{spacetimenergy}
\begin{split}
E &= J^z_0 + \tilde{J}^z_0 = (j + {k w \over 2}) + (j + Q+ {k w \over 2}), \\
&=2 j + k w  + P + {N + h_2 + h_{\rm int} \over w}, \\
L &= J^z_0 - \tilde{J}^z_0 = P + {N + h_2 + h_{\rm int} \over w}, \\
J_1 &= K^z_0 + \tilde{K}^z_0 = 2 j + P + k w ,\\
J_2 &= K^z_0 - \tilde{K}^z_0 = P .\\
\end{split}
\end{equation}
 
The degeneracy of states with a given value of $h_2, P$ and $h_{\rm int}$ is given to us by the partition functions for the $SU(2)$ WZW model and the internal CFT. There remains the issue of the degeneracy of,non spurious,  states with a given value of $N,Q$ that obey the physical state conditions \eqref{physicalarbitrary}. If we are interested only in the degeneracy and not in the actual construction of physical states, the formula for the spacetime partition function in the next section tells us to proceed as follows:
\begin{enumerate}
\item
Consider the affine primary of ${\hat{{\cal D}}}^{j}$ and act on it with the oscillator modes $J^+_{n \leq 0}, J^-_{n < 0}$, never acting with $J^z_{n<0}$. Let us call this set ${\cal Z}$. Now, consider the states obtained by spectral flow of the states in ${\cal Z}$ by $w$ units. Call this set ${\cal Z}^w$.
\item
 Consider the states in the tensor product ${\cal Z}^w \times {{\cal L}}^{\bar{j}({w})} \times H^0({\cal M}_{p,q})$. Decompose the character of this tensor product into representations of the Virasoro algebra \cite{Barabanschikov:2005ri}, 
pick out Virasoro primaries and impose the mass shell condition $L_0 \sim 0$. 
\end{enumerate}

The procedure above is valid for all states that lie in discrete representations and we have argued this is true for almost all assignments of charges.  We now turn to the case where $\tilde{J}^z_0 = {E-L \over 2} = {k w \over 2}$. As we explained in section \ref{review}, at this value of the charge, the giant graviton solution  can go off
to infinity. This infinite volume factor means that, upon quantization, the probability that a probe brane with these charges will be found at any finite value of $\rho$ is infitesimally small. Hence, to quantize solutions that have this value of $E-L$ we may simplify the formulae of Table \ref{global} by taking the $\rho \rightarrow \infty$ limit. Furthermore, since at infinity, such a solution 
must wrap the $\theta$ direction to have finite energy, we set $\theta' = w$. The remaining dynamical variables are $\rho, \zeta, \phi_1, \phi_2, z^a$ and we have:
\begin{equation}
\begin{split}
P_{\rho} &= -{k \over 2 \pi} \rho',~~ P_{\zeta}= {-k \over 2 \pi} \zeta',~~P_{z^{a}} = {-k \over 2 \pi}\left[g^{int}_{a b} z^{b'}\right] , \\
\tilde{P}_{\phi_1}&={k \over 2 \pi} \left[-\cos^2(\zeta)(\phi_1' - w/2) + \sin^2(\zeta) (\phi_2' - w/2) + {w \over 2} \right] = {k w \over 2 \pi} - \tilde{P}_{\phi_2} .
\end{split}
\end{equation}
All the complicated couplings between the different degrees of freedom have vanished in the $\rho \rightarrow \infty$ limit! Quantizing the $z^{a}$ and their conjugate momenta yields as we explained in Section \ref{review} to the left-moving sector of the non-linear sigma model on ${\cal M}_{p,q}$. Quantizing $\zeta, \phi_1, \phi_2$ leads as one may expect to the left moving sector of the $SU(2)$ WZW model at level $k$. The $\rho$ theory gives rise to a $U(1)$ theory. In terms of this $U(1)\times SU(2)$ theory, the spacetime energy and angular moment are given by:
\begin{equation}
\label{continousenergy}
E = k w + {N_{\rho} + h_2 + h_{\rm int} \over w},~~~ L = {N_{\rho} + h_2 + h_{\rm int} \over w} .
\end{equation}
where $h_2, h_{\rm int}$ are as above and $N_{\rho}$ is the level of the $U(1)$ theory. 

In fact this $U(1) \times SU(2)$ theory is nothing but the theory of long-strings studied in \cite{Seiberg:1999xz}.\begin{footnote}{A closely related theory was studied in \cite{Callan:1991at,Callan:1991dj,Callan:1991ky}}\end{footnote} We refer the reader to that paper for details but recount two salient conclusions. First, the $U(1) \times SU(2)$  theory admits a $N=4$ supersymmetric extension To obtain this we need to
improve the $U(1)$ model with a linear dilaton term that 
increases the central charge of the supersymmetric $U(1) \times SU(2)$ model to $6 (k - p q)$.  Now, with the $N=4$ supersymmetric sigma model on ${\cal M}_{p,q}$ we have a $N=4$ superconformal theory on the worldsheet and It is not hard to show from here that the entire superconformal symmetry carries over from the worldsheet to spacetime via \eqref{continousenergy}.

We also notice that the $N=4$ theory in the NS sector we have obtained above may be obtained by performing spectral flow in {\em spacetime} (to be distinguished from spectral flow on the worldsheet, that we have been discussing), on the theory of long-strings in the background of the zero mass BTZ black hole (the Poincare patch of $AdS_3$ with a circle identification) that was discussed in \cite{Mandal}. However, the quantization of strings in that background did not yield any of the discrete states that we have found in global AdS. This provides further evidence for the argument made in \cite{Lunin:2002bj} that the Poincare patch is {\em not} the correct background dual to the Ramond sector of the boundary theory.

We can also obtain the energy formula above using the analysis of states in continuous representations in \cite{Maldacena:2000hw} and this analysis shows us that they lie at the bottom of a continuum. 
The measure for continuous representations was worked out in \cite{Maldacena:2000kv} and since the supersymmetric states above correspond to a particular point (the bottom) and not to a range in the continuum, they are actually of measure zero.

Nevertheless, semi-classically we seem to have a complete story. For generic charges, ${1 \over 4}$ BPS states occur in discrete representations with an energy given by \eqref{spacetimenergy} and at special values of the charges, where the classical solutions can escape to infinity, they appear at the bottom of a continuum with energy given by \eqref{continousenergy}. 
In the exact analysis of the D-string carried out in Section \ref{spacetimepartfun}, this story is almost completely borne out except for the puzzling fact
that the measure for continuous representations vanishes in a  
neighbourhood of the point where we expect to find supersymmetric states. 
This leads to missing chiral primaries at special values of charges. We discuss
this issue and the implications of the observation above for the elliptic genus
in the next section.

\subsubsection{Half-BPS States}
We have provided a semi-classical description of ${1 \over 4}$ BPS states above. We now discuss ${1 \over 2}$ BPS states in spacetime. These are of the form $|{\rm chiral~primary} \rangle|{\rm chiral~primary} \rangle$. We will denote them by $|j_L, j_R  \rangle$ where $j_L, j_R$ are the R-charge values on the left and the right. 

It is easy to construct such states on the worldsheet. They are merely, states of the form $|{c}^{j,w}  \rangle |c^{j,w}  \rangle$. For concreteness, consider a D-string so that $k = Q_5$. The discussion above tells us that semi-classically, we should expect a chain of such states. First, we consider $w=0$ and all possible values of $j$. This leads to chiral states in spacetime of the form $|j,j \rangle$ with values of ${1 \over 2} \leq j \leq {Q_5 - 1 \over 2}$. Now, we spectral flow these states to obtain states of the form $|j + {Q_5 w \over 2}, j + {Q_5 w \over 2} \rangle$. There are gaps in the chiral primary spectrum at $j = {Q_5 w \over 2}$ because at these values the chiral-primaries lie in the continuum as we discussed above. The `exclusion principle' tells us that we must restrict to $w \leq Q_1$. 
This semi-classical picture captures all the essential features of the exact analysis that we perform in the next section. The inclusion of fermionic zero-modes gives a degeneracy to each element of this chain. Furthermore, it is also possible to have $j_{L}=j_{R} \pm {1 \over 2}$. The exact spectrum is worked out in the next section.

Somewhat more curiously, we seem to get a copy of this series of ${1 \over 2}$ BPS states states from each kind of $(p,q)$ probe.  However, this is not a surprise when we recall that the probe solutions corresponding to chiral-chiral primaries are geodesics that do not know anything of the internal structure of the probe. Hence, to obtain the correct spectrum of half-BPS states on the boundary, we should count the chiral-primaries only once and not repeatedly. The simplest way to avoid over-counting is to consider the chiral primaries obtained from the single and multi-particle states of the D-string. In the next section, we show how the ${1 \over 2}$ BPS spectrum of the boundary theory may be reproduced this way. We could also use a different probe although chiral primaries obtained from a multi-particle state of the D-string may be the same as chiral-primaries obtained from a single particle state of a more complicated probe.

\section{Exact Analysis of the D-string}
\label{spacetimepartfun}
In this section, we will analyze the exact spacetime partition function for the D-string. When the string coupling is large in the D-brane picture, $Q_1 >> {v Q_5 \over g^2}$, we can perform a S-duality to obtain a weakly coupled F-NS5 system. The motion of D-strings in global $AdS_3 \times S^3$ with RR fluxes but no
NS fluxes is dual to the propagation of F-strings in $AdS_3 \times S^3$ with NS fluxes but no RR fluxes. This system has been widely studied. For some early studies of string propagation on $AdS_3$ and its relation to the AdS/CFT correspondence, see \cite{Giveon:1998ns,Kutasov:1999xu,deBoer:1998pp}. In this section, we will rely heavily on the papers \cite{Maldacena:2000hw,Maldacena:2000kv}. Please also refer to these papers for a review of the early literature on string theory on $AdS_3$. For later studies, see \cite{Giveon:2003ku,Pakman:2003cu,Pakman:2003kh}. The supersymmetric extension of the partition function of \cite{Maldacena:2000kv} that we will use here was studied in \cite{Israel:2003ry}.

For ease of presentation,  we will work in the background $AdS_3 \times S^3 \times T^4$.  The calculations we perform here may be easily repeated for $K3$, and none of the results we will obtain here are affected. Our plan of attack will be to generalize the spacetime partition function of the bosonic string calculated in \cite{Maldacena:2000kv} to the superstring. By taking various limits of this partition function, we will then obtain expressions for the degeneracies of ${1 \over 4}$ BPS states and ${1 \over 2}$ BPS states. We will also discuss the elliptic genus. This section contains several messy manipulations with infinite products that we relegate to Appendix \ref{technicaldetailspartfun} leaving only the results for the main text.

We start with the formula for the spacetime partition function of the bosonic string, derived in  \cite{Maldacena:2000kv}. In terms of the $AdS$ energy, $E$, and angular momentum $L$, the levels of the boundary CFT are given by \cite{Aharony:1999ti}
\begin{equation}
L_0^b = {E + L \over 2} ,~~ \tilde{L}_0^b = {E-L \over 2} .
\end{equation}
Then
\begin{equation}
\label{spacetimepart}
\begin{split}
Z(\beta, \bar{\beta}) &= {\rm Tr}_{\rm single-particles} e^{-\beta L_0^b - \bar{\beta} \tilde{L}_0^b}\\
&= {b (Q_5 - 2)^{1/2} \over 8 \pi} \int_0^{\infty} {d \tau_2 \over \tau_2^{3/2}} \int_{-1/2}^{1/2} d \tau_1 e^{4 \pi \tau_2(1 - {1 \over 4(Q_5-2)})} Z_{\rm SU(2)}(q,\bar{q}) Z_{\rm int}(q,\bar{q}) \\
&\times {e ^{- (Q_5 - 2) b^2/{4 \pi \tau_2}} \over |\sinh({\beta}/2)|^2} \left| \prod_{n=1}^{\infty} {1 - q^n \over (1 - e^{\hat{\beta}} q^n)(1 - e^{{-\beta}} q^n)} \right|^2, \\
\end{split}
\end{equation}
where, $q = e^{2 \pi i \tau}, \tau = \tau_1 + i \tau_2$. The formula above is valid when $b \equiv {\rm Re}(\beta) = {\rm Re}(\bar{\beta}) > 0$. Here, $Z_{\rm int/SU(2)} = Tr_{T^4/SU(2)} (q^{L_0^b} \bar{q}^{\tilde{L}_0^b})$. Notice the absence of the usual ${c \over 24}$ shift. This zero point energy has already been taken into account in \eqref{spacetimepart}.

Let us expand some of the terms in the formula above. First, we consider the $SU(2)$ partition function. We will denote the level of this model by $k-2$. At this level, the character of the representation ${\cal L}^j$, built on an affine primary with weight $j$ is given by\cite{DiFrancesco:1997nk}:
\begin{equation}
\label{sutwochar}
\chi_j^{k-2}(\tau, \rho) = tr(q^{L_0^{\rm SU(2)}} z^{K_0^z})= { q^{{1 \over 8} - {1 \over 4 k}} \sum_{n \in \mathbb{Z}} q^{(j + {1 \over 2} + k n)^2 \over k} (z^{j + {1 \over 2} + k n} - z^{-(j + {1 \over 2} + k n)}) \over i \theta_1(\rho, \tau)},
\end{equation}
where, $z = e^{2 \pi i \rho}$. 

Naively, one may think that the formula above has a pole of order $1$ when $z = q^{w}$ where $w \in \mathbb{Z}$. However, this is not the case, because the numerator also vanishes for this assignment of chemical potentials. The formula above is valid whenever $|q| < 1$. 
The partition function of the bosonic $SU(2)$ WZW model, at level $k-2$ is given by
\begin{equation}
Z_{SU(2)- bosonic}^{k-2}(\tau,\rho) = \sum_{j = 0}^{k/2 - 1} |\chi_j^{k-2}(\tau,\rho)|^2 .
\end{equation}
The spacetime $SU(2)$ charges are measured by the worldsheet $SU(2)$ charges, and in the notation of Table \ref{global}, the spacetime $SU(2)$ angular momenta $J_L, J_R$ are given by
\begin{equation}
J_R = {J_1 + J_2 \over 2} = \tilde{K}^z_0 ,~~ J_L = {J_1 - J_2 \over 2} = {K}^z_0.
\end{equation}

Next, we need to include fermions and generalize the expression to the superstring. As explained in \cite{Giveon:1998ns} and references therein, the addition of fermions in the $SL(2,R)$ and $SU(2)$ WZW models is simple.  We obtain decoupled fermionic and bosonic WZW models, except that the level of the bosonic $SL(2,R)$ model is shifted to $Q_5 + 2$ and the level of the bosonic $SU(2)$ model is shifted to $Q_5 - 2$. To generalize expression \eqref{spacetimepart} then, we merely need to add in the worldsheet partition function for the fermions and alter the zero-point energies and levels appropriately. This is done in Appendix \ref{technicaldetailspartfun} which also discusses the sum over R-NS sectors and the GSO projection. The final ingredient we need is the worldsheet partition function of the internal $T^4$.

Putting all of this together we find that the full partition function of a F-string propagating in $AdS_3 \times S^3 \times T^4$ with $Q_5$ units of NS flux is given by
\begin{equation}
\label{susypart}
\begin{split}
&Z(\beta, \bar{\beta}, \rho, \bar{\rho}) = {\rm Tr}_{\rm single-particles} e^{-\beta L_0^b -\bar{\beta} \tilde{L}_0^b - \rho J_L - \bar{\rho} J_R} \\
&= {b Q_5^{1 \over 2} \over 2  \pi} \int {d \tau_2  \over \tau_2^{3 \over 2}} \int_{-1 \over 2}^{1 \over 2} d \tau_1  \left[ e^{- Q_5 {b^2 \over 4 \pi \tau_2}  \over \tau_2} 
\sum_{j = 0}^{{Q_5 \over 2} - 1}\left|\left({\sum_{n \in \mathbb{Z}} q^{(j + {1 \over 2} + Q_5 n)^2 \over Q_5} (z^{j + {1 \over 2} + Q_5 n} - z^{-(j + {1 \over 2} + Q_5 n)}) \over \theta_1({i \rho \over 2 \pi}, \tau)} \right)\right|^2 \right. \\
&\times \left. \left( \sum_{\Gamma^{4,4}} q^{p_L^2} \bar{q}^{p_R^2}\right)  \left|{\left(\theta_2(i {\beta + \rho \over 4 \pi}) \theta_2(i {\beta - \rho \over 4 \pi})\right)^2 \over \theta_1({i \beta \over 2 \pi}, \tau) \eta(\tau)^6} \right|^2 \right] .
\end{split}
\end{equation}
The theta functions can be decoded by looking in Appendix \ref{thetaappendix}. The contribution of  the zero mode momenta and winding on the $T^4$ is contained in the sum over the lattice $\Gamma^{4,4}$. However, as in \cite{Maldacena:1999bp}, we will focus on states that carry $0$ charge under $p_L$ and $p_R$.  

Some of the symmetry of the spacetime theory is already visible in \eqref{susypart}. The zero modes of the $\theta$ functions generate the global supergroup $SU(1,1|2)$. The zero modes of the four theta functions in the numerator correspond to the action of $8$ left moving and $8$ right moving supercharges. The zero modes of the theta functions in the denominator correspond to the action of $K^{-}_0$ and $J^{+}_0$. 

It is not hard to repeat the calculation of the partition function above for $K3$; if we work at a value of $K3$ moduli where $K3$ is just $T^4/Z_2$, then we merely need to modify the $T^4$ partition function above by adding in twisted sectors and projecting onto invariant states. This will add 3 more terms to the last line of \eqref{susypart}, lead to a different spectrum of chiral primaries below and give a finite result for the elliptic genus. However, none of our conclusions or puzzles below are affected.

\subsection{${1 \over 4}$ BPS Partition Function}
\label{quarterbpspartfun}
The ${1 \over 4}$ BPS partition function for the D-string is obtained from the formula \eqref{susypart} by taking the limit $\bar{\beta} \rightarrow \infty, \bar{\rho} \rightarrow -\infty$ keeping $\bar{\beta} + \bar{\rho} = -\mu$ finite. It is shown in Appendix \ref{technicaldetailspartfun} that in this limit we can ignore all right-moving oscillator contributions. It is then possible to do the integral over $\tau_2$ in \eqref{susypart} and the remaining integral over $\tau_1$ then just provides a level matching condition. 

Let us define the function $f$ by:
\begin{equation}
\label{fdef}
{\theta_2({u - v \over 2}, \tau)^2 \theta_2 ({u + v \over 2},\tau)^2 \over -i \theta_1(u,\tau) \eta(\tau)^6} \chi_{j}^{Q_5-2}(\tau,v)  = \sum_{Q,P,h} f_j(Q,P,h) e^{2 \pi i u Q} e^{2 \pi i v (P + j + {1 \over 2})} 
e^{2 \pi i \tau  \{{j(j+1) \over Q_5} + h\}} ,
\end{equation}
where we expand the left hand side in the regime where $0 < {\rm Im}(u) < {\rm Im}(\tau)$. Note, that at any given power of $q = e^{2 \pi i \tau}$, the expansion in powers of $z=e^{2 \pi i v}$ terminates after a finite number of terms. 
 In terms of this function $f$, the ${1 \over 4}$ BPS partition function is given by a remarkably
simple formula.
\begin{equation}
\label{quarterbpspart}
\begin{split}
Z_{1 \over 4}(\beta, \rho, \bar{\mu}) &= 4 \cosh^2{\bar{\mu} \over 4} \sum_{w \geq 0} \sum_{j = 0}^{{Q_5 \over 2} - 1} e^{-\bar{\mu} (j + {Q_5 w \over 2} + {1 \over 2})} \\  &\times \sum_{\tiny {\begin{array}{l}w (Q - P) \\ - h  = 0 \end{array}}} f_j(Q , P, h) e^{-\beta (Q + j + {1 \over 2} + {Q_5 w \over 2}) - \rho(P+j + {1 \over 2} + {Q_5 w \over 2})}  .
\end{split}
\end{equation}
Comparing this with \eqref{spacetimenergy}, if we redefine $j \rightarrow j + {1 \over 2}$ we find exact agreement with the formulae for the charges given there. Thus we see, as promised, that the semi-classical formula in section \ref{semiclassicalsection} has given us an exact answer with all factors of $1$ correct, at least for the D-string. It is tempting to conjecture that this is also the case for $(p,q)$ strings. 

Note that, for $w > 0$, we may replace $Q$ in the second line by $Q = P + {h \over w}$. For $w = 0$, the sum runs over terms that have $h = 0$. These terms
come from the zero modes in the theta functions in \eqref{fdef} and give us 
the graviton multiplets described in \cite{deBoer:1998us}.

Although we have written the sum \eqref{quarterbpspart} over all positive $w$, the exclusion principle proposed in \cite{Argurio:2000tb} along the lines of \cite{Maldacena:1998bw,deBoer:1998us} instructs us to cut off this sum at $w = Q_1$.

\subsection{${1 \over 2}$ BPS Partition Function}
\label{halfbpspartfun}
Now we will try and obtain the spectrum of chiral-chiral states. To do this, in addition to the limit above,  we need to take the limit $\beta \rightarrow \infty$, $\rho \rightarrow -\infty$, keeping $\beta + \rho = -\mu$ finite. It is shown in Appendix \ref{technicaldetailspartfun} that in this limit, we can ignore all contributions from the theta functions except for the zero modes of 
$\theta_2(i {\beta + \rho \over 4 \pi})$ in \eqref{susypart}.
The character for the chiral primaries then becomes:
\begin{equation}
\label{chiralprim}
\begin{split}
&Z_{1 \over 2}(\mu, \bar{\mu}) = {\rm tr}_{\rm chiral-primaries} e^{\mu K^z_0 + \bar{\mu} \tilde{K^z_0}} \\
&= \lim {b Q_5^{1 \over 2} \over 2 \pi}  \int {d \tau_2  \over \tau_2^{3 \over 2}} \int_{-1 \over 2}^{1 \over 2} d \tau_1  e^{- Q_5 b^2  \over 4 \pi \tau_2} \sum_{j=0}^{Q_5/2 - 1} \left|q^{(j + {1 \over 2} + Q_5 n)^2/Q_5}(z^{j + {1 \over 2} + Q_5 n} - z^{-(j + {1 \over 2} + Q_5 n)} (2 \cosh^2{\mu \over 4}) \right|^2 \\
&=\sum_{n \in \mathbb{Z}^+} \sum_{j = 0}^{Q_5/2 - 1}  \left(2\left( t^{1 \over 2} + {1 \over t^{1 \over 2}} + \bar{t}^{1 \over 2} + {1 \over \bar{t}^{1 \over 2}}\right) + 4 + {t^{1\over 2} \over \bar{t}^{1 \over 2}} + {\bar{t}^{1 \over 2} \over t^{1 \over 2}} + t^{1\over 2} \bar{t}^{1 \over 2} + {1 \over t^{1 \over 2} \bar{t}^{1 \over 2}}\right)(t \bar{t}) ^{j + {Q_5 n \over 2} + {1 \over 2}} ,
\end{split}
\end{equation}
where $t = e^{\mu}$.
This is in agreement with \cite{Kutasov:1998zh}. 
This analysis can easily be repeated for $K3$ to obtain a spectrum in agreement with \cite{deBoer:1998ip,deBoer:1998us}. If we apply the exclusion principle mentioned above, then the highest power of $t$ that appears above is ${Q_1 Q_5 \over 2}$.

 Notice, though that the chiral primaries that correspond to $j = Q_5/2$ in the series above are not present. We expect this from our semi-classical analysis above. On the boundary, these missing chiral primaries result from the small
instanton singularity \cite{Seiberg:1999xz}; in the bulk this phenomenon 
was first noticed in \cite{Argurio:2000tb}. There, it was suggested, as we reasoned above,  that these missing chiral primaries disappear into the continuum.

Let us examine this hypothesis.  In Appendix \ref{technicaldetailspartfun} we show that chiral primaries can occur in the continuous spectrum if the condition
\begin{equation}
\label{continuousenergy}
j + Q_5 n + {1 \over 2} = {Q_5 w \over 4} + {1 \over w} ({s^2 \over Q_5} + {(j + Q_5 n + {1 \over 2})^2 \over Q_5}) 
\end{equation}
is met with $w$ being some integer. This can only happen if:
\begin{equation}
\begin{split}
s &= 0 , \\
j + Q_5 n + {1 \over 2} &= {Q_5 w \over 2} .
\end{split}
\end{equation}
This appears promising before we realize that this condition cannot be met because the sum over $j$ runs from $0 \ldots {Q_5 \over 2} - 1$. We discuss this issue further in Section \ref{comparisonsymmetric}.

\subsection{Elliptic Genus}
\label{ellipticgenussection}
We now turn to a study of the elliptic genus. The elliptic genus is defined as
\begin{equation}
E(\beta, \rho) = {\rm tr}\{e^{-\beta L_0^b - \rho J_L -\bar{\beta}(\tilde{L}_0^b - J_R)} (-1)^{2 J_R}\} .
\end{equation}
The chemical potential $\bar{\beta}$ is purely formal; the elliptic genus is independent of this parameter.

For $T^4$ the elliptic genus vanishes due to fermion zero modes. Although, we could repeat this calculation for $K3$ to obtain a finite elliptic genus, we will instead consider the quantity:
\begin{equation}
\label{modifiedindex}
E_2(\beta, \rho) = {\rm tr}\{e^{-\beta L_0^b - \rho J_L -\bar{\beta}(\tilde{L}_0^b - J_R)} (-1)^{2 J_R} J_R^2\} .
\end{equation}
This quantity was defined and studied in \cite{Maldacena:1999bp}, specifically to study BPS states in toroidal string theory.  The trace is taken only over states that have no $U(1)^4$ charge. In the formula of $\eqref{susypart}$ this instructs us to drop the sum over $\Gamma^{4,4}$. 
We then find:
\begin{equation}
\label{ellipticsimple}
\begin{split}
&E_2(\beta, \rho) = \left.{\partial^2 Z_{1 \over 4}(\beta, \rho, \bar{\mu}) \over \partial \bar{\mu}^2}\right|_{\bar{\mu} = 2 \pi i}\\
&= {1 \over 2} \sum_{w \geq 0} \sum_{j = 0}^{{Q_5 \over 2} - 1} \sum_{\tiny {\begin{array}{l}w (Q - P) \\ - h  = 0 \end{array}}} (-1)^{2 j} f_j(Q , P, h) e^{-\beta (Q + j + {1 \over 2} + {Q_5 w \over 2}) - \rho(P+j + {1 \over 2} + {Q_5 w \over 2})}  .
\end{split}
\end{equation}
Notice, that several cancellations occur in the expression above because of the term $(-1)^{2 j}$ above. 

\subsection{Comparison to the Symmetric Product}
\label{comparisonsymmetric}
Before we compare our results for the elliptic genus and the ${1 \over 2}$ BPS
partition function to the symmetric product, let us briefly review some known results.  In \cite{deBoer:1998ip,deBoer:1998us}, de Boer found the spectrum of gravitons in
 $AdS_3 \times S^3 \times K3$ and organized it into short representations of the relevant AdS supergroup $SU(1,1|2)_L \times SU(1,1|2)_R$. His results may be generalized to $T^4$, and in that case
the spectrum of single-particle gravitons described above consists of the ${1 \over 2}$ BPS states of Section \eqref{halfbpspartfun}  and their descendants under the generators of this global supergroup. In formula \eqref{susypart}, the action of these global generators is seen in the zero-modes of the theta functions. Now, two results were obtained in \cite{deBoer:1998us} (See, also \cite{Kraus:2006nb}). First, it was found that the spectrum of chiral-chiral primaries of the symmetric product up to energies ${Q_1 Q_5 \over 2}$ could be found by multi-particling the spectrum of single particle chiral-chiral primaries of supergravity subject to a suitable exclusion principle. Second, with an extension 
of this exclusion principle, it was found that the elliptic genus of supergravity also agreed with the elliptic genus of the symmetric product till the energy ${Q_1 Q_5 \over 4}$. For the case of $T^4$ a similar result regarding the modified index \eqref{modifiedindex} was proved in \cite{Maldacena:1999bp}.

These results are surprising, because naively one would expect supergravity to 
be valid till an energy $Q_5$ (assuming $Q_5 < Q_1$), and expect stringy 
effects to take over beyond that. Indeed, from formula \eqref{quarterbpspart}, we see
that the ${1 \over 4}$ BPS spectrum of the string theory agrees with supergravity 
till energies of order $Q_5$ (i.e in the zero-winding sector) but disagrees 
for energies larger than that. 

However, the result of section \ref{halfbpspartfun}  shows, as was expected from the semi-classical analysis of Section \ref{chiralchiralsection},  that the ${1 \over 2}$ BPS spectrum of the full string theory agrees with the ${1 \over 2}$ BPS spectrum of supergravity up to an energy ${Q_1 Q_5 \over 2}$, barring some missing chiral-primaries. Modulo this complication, the calculation of \cite{deBoer:1998us} shows us that multi-particling the spectrum of Equation \eqref{chiralprim} with an appropriate exclusion principle at high-energies will reproduce the spectrum of chiral-chiral states of the symmetric product.

The issue of missing chiral-primaries acquires greater urgency in a consideration
of the elliptic genus.\begin{footnote}{Here, we are tacitly assuming that we are on $K3$. For $T^4$ where the elliptic genus vanishes, everything in our discussion is valid with ``elliptic genus'' replaced by the modified index \eqref{modifiedindex}}\end{footnote} From \eqref{ellipticsimple}, we see that for left moving conformal
weight larger than ${Q_5 \over 2}$, the elliptic genus contains contributions 
from ${1 \over 4}$ BPS states that are not seen in supergravity. Hence, multi-particling this spectrum leads to a mismatch with the elliptic genus of the 
symmetric product. This, however,  does not contradict any theorem because 
as we have mentioned the boundary theory is singular on this submanifold of 
moduli space and has a continuum in its spectrum; this invalidates the 
index theorems that protect the elliptic genus \cite{Cecotti:1992qh}.

By modular invariance, the high temperature behaviour of the elliptic genus is dominated by the
lowest energy supersymmetric states in the spectrum. Since these new ${1 \over 4}$ BPS 
contributions appear after an energy gap, their effect on the high temperature behaviour
is {\em exponentially subleading}. So, they do not affect entropy counting calculations. However, it would be interesting to understand the physical 
interpretation of these subleading contributions in the spirit of \cite{Dijkgraaf:2000fq}. An interesting possibility is that these subleading terms correspond
to multi black holes. 

As we deform the theory away from this point in moduli space, the continuum 
must resolve to give rise to new ${1 \over 4}$ BPS states $|{\rm anything} \rangle |{\rm chiral~primary} \rangle$ with chiral primaries 
corresponding to $j = {Q_5  \over 2}$ in the sum \eqref{chiralprim}. This is necessary to supply the missing ${1 \over 2}$ BPS states and the right ${1 \over 4}$ BPS states to cancel the extra terms in \eqref{ellipticsimple}. Schematically, this happens as follows. 

On this submanifold of moduli space, the single partition function of string theory may be written as 
\begin{equation}
\label{schematic} 
Z(\beta,\bar{\beta}, \rho, \bar{\rho}) = \sum_{h, \bar{h}, r, \bar{r}} n(h, \bar{h}, r, \bar{r}) e^{-\beta h - \bar{\beta} \bar{h} - \rho r - \bar{\rho} {\bar r} }  + 
\sum_{r, \bar{r}} \int {\cal \rho}(h, \bar{h}, r, \bar{r})e^{-\beta h - \bar{\beta} \bar{h} - \rho r - \bar{\rho} {\bar r} } \, d h d \bar{h},
\end{equation}
which represents the contributions from both the discrete and continuous representations. 
We have seen that the second term does not contribute to the ${1 \over 4}$ BPS partition function
because
\begin{equation}
\rho(h, \bar{h}, r, \bar{h}) = 0, \forall h, \bar{h}, r .
\end{equation}
Now, the energy formula \eqref{continuousenergy} does allow states with $\bar{r} = \bar{h}$ to 
exist in continuous representations. The reason the measure above vanishes for these states 
is that in the $SU(2)$ WZW model at level $Q_5 - 2$, there is no lowest weight 
representation of weight ${Q_5 \over 2} - {1 \over 2}$. In fact, from formula \eqref{sutwochar}, we see that
\begin{equation}
\label{allnull}
\chi^{Q_5 - 2}_{{Q_5 \over 2} - {1 \over 2}}(\tau, \rho) = 0, \forall \tau, \rho .
\end{equation}
The character of a representation may be obtained by  symmetrizing the character of the 
corresponding Verma module over the Weyl group to remove null states \cite{fuchs1997sla}. 
So, loosely speaking
we can interpret \eqref{allnull} to mean that all states in this representation are null.

As we deform the theory away from this point in moduli space, we can imagine the supersymmetric
spectrum changing via a two step process.
In the first step, the continuum resolves into discrete states
\begin{equation}
\label{resolved}
\sum_{r, \bar{r}} \int {\cal \rho}(h, \bar{h}, r, \bar{r})e^{-\beta h - \bar{\beta} \bar{h} - \rho r - \bar{\rho} {\bar r}} \, d h d \bar{h} \rightarrow  \sum_{h, \bar{h}, r, \bar{r}} n'(h, \bar{h}, r, \bar{r}) e^{-\beta h - \bar{\beta} \bar{h} - \rho r - \bar{\rho} {\bar r} }.
\end{equation}
And in the second step ${1 \over 4}$ BPS discrete states combine into long representations leaving behind a reduced supersymmetric spectrum.  

However, as we move away from this point in moduli space by turning on RR fields, we 
also deform the worldsheet current 
algebra. Under such a deformation, the RHS of \eqref{allnull} may jump from zero. Then, \eqref{continuousenergy} tells us that it is possible that
\begin{equation}
n(h, \bar{h}, r, \bar{h}) \neq 0,~{\rm for}~\bar{h} \in \{ {Q_5 w \over 2}, {Q_5 w \over 2} \pm {1 \over 2} \},
\end{equation}
where $w$ is a positive integer.
These new discrete states could provide the missing chiral-primaries and also pair up with 
the extra ${1 \over 4}$ BPS states to remove them from the supersymmetric spectrum. It would be
nice to have a more quantitative understanding of this process.

\subsection{Higher Probes}
The partition function for the entire theory is obtained by summing, not only over states of the D-string but also over the more complicated $(p,q)$ probes. 
Now, if we take the action \eqref{Polyakov} with the substitutions \eqref{substitutions} seriously, and attempt to quantize it like the fundamental string, we are left with a theory that, for a generic $(p,q)$ probe, has too large a central charge. This is not a surprise, because the manipulations that led to \eqref{substitutions} were classical in nature. A bona-fide analysis of supersymmetric states in these higher probes must start with the worldvolume theory of the D5 brane. 

However, the semi-classical analysis of Section \ref{review} and the analysis of long-strings in Section \ref{semiclassicalsection} suggest a possible resolution. In formulae \eqref{semiclassicaldirectwo}, \eqref{directsemiclassical}, \eqref{continousenergy} the non-linear sigma model on ${\cal M}_{p,q}$ made its appearance. In the bosonic case, it seems possible to generalize the exact
analysis of the D-string by simply substituting the bosonic partition function of ${\cal M}_{p,q}$ in place of $Z_{\rm int}$ in formula \eqref{spacetimepart}, without changing the zero-point energy (the coefficient of $\tau_2$ in the exponent) at all.

To understand this better, consider the following analogy. Say, we are trying to quantize a bosonic string in $d$ dimensions, where $d$ is not necessarily $26$. Let us choose light cone gauge, and impose the mass-shell condition $(L_0 - 1)|\Omega \rangle = 0$. This leads to a spectrum that is free of the Lorentz anomaly. At the massless level, we obtain a representation of $SO(d-2)$ and at higher levels the spectrum reorganizes itself into representations of $SO(d-1)$. Of course, we cannot consistently introduce interactions in this theory, but if we are interested {\em only} in the spectrum, this procedure leads to a sensible result. 

For our case, the supersymmetric spectrum can perhaps be obtained by appropriately supersymmetrizing this bosonic spectrum obtained in this manner.

We conclude this section with a speculative possibility.  
It is possible that if we
sum the contributions to the elliptic genus over all the different $(p,q)$ probes, the contributions from 
all states except for ${1 \over 2}$ BPS states cancel. To check this possibility, however, we need to be able to exactly quantize the more complicated $(p,q)$ probes. This is a very interesting problem that we leave to future work.

\section{Results}
\label{resultsection}

In this paper, we first developed an alternative approach to classical probe brane solutions in global $AdS_3$, in terms of the  `Polyakov' action. We showed that the canonical structure on the space of ${1 \over 4}$ BPS brane probes found in \cite{Mandal} was the same as the canonical structure on the solutions \eqref{solveright} of the sigma-model \eqref{Polyakov} except that 
the `Polyakov' approach also allowed us to identify the classical solutions corresponding to ${1 \over 2}$ BPS states. We found that these states were described by geodesics that do not see `stringy' effects even at energies above $Q_1$ and $Q_5$. This explained several facts about the spectrum of ${1 \over 2}$ BPS states that had, hitherto, been puzzles.

 Second, the `Polyakov' approach allowed us to recast the problem of quantizing these 
supersymmetric  probes as a problem of quantizing the sigma model \eqref{Polyakov} and picking out the physical subsector of the Hilbert space. We followed this procedure and found that, generically,  the quantization of ${1 \over 4}$ BPS brane probes in global $AdS_3 \times S^3 \times K3/T4$ leads to states in discrete representations of the $SL(2,R)$ WZW model with energy, given as a function of charges, by \eqref{spacetimenergy}. Semi-classically, at special values of the charges, the ${1 \over 4}$ BPS states are at the bottom of a continuum. Quantizing these probes leads to the long strings studied by Seiberg and Witten with energy given as a function of charges by \eqref{continousenergy}. 

The presence of these discrete states in global AdS is in sharp contrast to the result obtained by quantizing ${1 \over 4}$ BPS brane probes in the background of the zero mass BTZ black hole (Poincare patch with a circle identification). There, we only obtain states at the bottom of a continuum. So our results here bolster the argument made in \cite{Lunin:2002bj} that the Poincare patch is not the correct background dual to the Ramond sector of the boundary theory. 

Since, the ${1 \over 4}$ BPS brane probe solutions cease to exist if we turn on the bulk NS-NS fields or theta angle, we concluded that this leads to a jump in the ${1 \over 4}$ BPS partition function. 

By exactly quantizing the D-string we verified the energy formula \eqref{spacetimenergy}. 
Furthermore, by taking the appropriate limit of the ${1 \over 4}$ BPS partition function we 
obtained, in equation \eqref{chiralprim}, the spectrum of single particle 
chiral-chiral primaries  of the D-string. Modulo the issue of some `missing' chiral primaries at special charges (that result from singularities of the boundary theory at this point in moduli space), multi-particling this spectrum reproduces the spectrum of chiral-chiral primaries of the symmetric product.
 In section \ref{ellipticgenussection}, we found that stringy ${1 \over 4}$ BPS states in discrete representations 
contribute to the 
bulk elliptic genus on the special submanifold of moduli space where the background 
NS-NS fluxes and theta angle are set to zero. This leads to subleading 
terms in the elliptic genus of the theory on this submanifold of 
moduli space that are not present in the elliptic genus of the symmetric product.
In Section \ref{comparisonsymmetric} we showed that as we move away from 
this special submanifold, the continuum must resolve in a specific way to 
cancel these additional contributions and supply the missing chiral primaries.

It would be of interest to extend our analysis of $(p,q)$ bound state probes beyond the semi-classical approximation. This is an important direction for future work.

\section*{Acknowledgements}
I would like to thank S. Minwalla for his advice throughout this project and G. Mandal and M. Smedback for collaboration in the early stages of this work. I would also like to thank  A. Dabholkar, J. de Boer, F. Denef, R. Gopakumar, L. Grant,  M. Guica, J. Maldacena, S. Mukhi, K. Narayan, D. Tong, S. Trivedi and especially S. Lahiri, S. Nampuri and K. Papadodimas for helpful discussions. 

\section*{Appendices}
\appendix
\section{Technical Details of the Spacetime Partition Function}
\label{technicaldetailspartfun}
In this appendix, we will fill in the details that lead to the results of section \ref{spacetimepartfun}.
\subsection{Partition function}
 To generalize the bosonic partition function \eqref{spacetimepart} we need to add in fermions and the $\beta \gamma$ ghosts,  sum over R-NS sectors, impose the GSO projection and explicitly include the partition function of $T^4$. 

First, consider the worldsheet partition function for the $SL(2,R), SU(2)$ and $T^4$ fermions and $\beta \gamma$ ghosts. For each of these, we can calculate the quantity:
\begin{equation}
\begin{split}
Z(a,b)(\beta, \rho, \tau) &= {\rm Tr}((-1)^{b F} e^{\rho K^z - \beta J^z + 2 \pi i \tau(L_0 - {c \over 24})} ,\\ 
\psi(\sigma + 2 \pi) &= (-1)^a \psi(\sigma) .
\end{split}
\end{equation}

These partition functions are listed explicitly in the Table below.
\begin{equation}
\begin{array}{|c|c|c|c|c|} 
\hline
& Z(0,0) & Z(1,0)&  Z(0,1)  &  Z(1,1)\\
\hline 
{\rm SL(2,R)~fermions }& {\theta_2({i \beta \over 2 \pi}, \tau) \theta_2(0,\tau)^{1 \over 2} \over \eta(\tau)^{3 \over 2}} & {\theta_3({i \beta \over 2 \pi}, \tau) \theta_3(0,\tau)^{1 \over 2} \over \eta(\tau)^{3 \over 2}} & {\theta_1({i \beta \over 2 \pi}, \tau) \theta_1(0,\tau)^{1 \over 2} \over \eta(\tau)^{3 \over 2}} & {\theta_4({i \beta \over 2 \pi}, \tau) \theta_4(0,\tau)^{1 \over 2} \over \eta(\tau)^{3 \over 2}}\\
\hline
{\rm SU(2)~fermions }& {\theta_2({i \rho \over 2 \pi}, \tau) \theta_2(0,\tau)^{1 \over 2} \over \eta(\tau)^{3 \over 2}} & {\theta_3({i \rho \over 2 \pi}, \tau) \theta_3(0,\tau)^{1 \over 2} \over \eta(\tau)^{3 \over 2}} & {\theta_1({i \rho \over 2 \pi}, \tau) \theta_1(0,\tau)^{1 \over 2} \over \eta(\tau)^{3 \over 2}} & {\theta_4({i \rho \over 2 \pi}, \tau) \theta_4(0,\tau)^{1 \over 2} \over \eta(\tau)^{3 \over 2}}\\
\hline
{\rm T^4~fermions}& {\theta_2(0, \tau)^2 \over \eta(\tau)^2} & {\theta_3(0, \tau)^2 \over \eta(\tau)^2}& {\theta_1(0, \tau)^2 \over \eta(\tau)^2}& {\theta_4(0, \tau)^2 \over \eta(\tau)^2} \\
\hline
{\rm \beta \gamma~ghosts}& {\theta_2(0, \tau) \over \eta(\tau)} & {\theta_3(0, \tau) \over \eta(\tau)}& {\theta_1(0, \tau) \over \eta(\tau)}& {\theta_4(0, \tau) \over \eta(\tau)} \\
\hline
\end{array}
\end{equation}

Finally, the worldsheet fermionic partition function may be written as
\begin{equation}
\begin{split}
&Z_{\rm fer}(\beta, \rho, \bar{\beta}, \bar{\rho}, \tau, \bar{\tau}) \\&= \left|{\theta_2({i \beta \over 2 \pi}) \theta_2({i \rho \over 2 \pi}) \theta_2(0)^2 - \theta_1({i \beta \over 2 \pi}) \theta_1({i \rho \over 2 \pi}) \theta_1(0)^2 + \theta_4({i \beta \over 2 \pi}) \theta_4({i \rho \over 2 \pi}) \theta_4(0)^2 -\theta_3({i \beta \over 2 \pi}) \theta_3({i \rho \over 2 \pi}) \theta_3(0)^2 \over \eta(\tau)^6} \right|^2\\
&= \left|{\theta_2 (i {\beta + \rho \over 4 \pi})^2 \theta_2(i { \beta - \rho \over 4 \pi}) ^2 \over \eta(\tau)^6} \right|^2 .
\end{split}
\end{equation}
where, in the last step, we have used the Riemann identity. We can think of this as passing from the R-NS formalism to the Green Schwarz formalism. 

\subsection{The Integral}
\subsubsection{Chiral Primaries}
Recall, as explained in \cite{Maldacena:2000kv} that the integral in \eqref{susypart} starts by writing
\begin{equation}
e^{-k \beta^2 \over 4 \pi \tau_2} = {- 8 \pi i  \over \beta} \left({\tau_2 \over k}\right)^{3 \over 2} \int_{-\infty}^{\infty} dc \, c e^{{- 4 \pi \tau_2 \over k} c^2 + 2 i \beta c} .
\end{equation}
Now, notice that if we expand the other $\theta$ functions in \eqref{susypart} then, we will get an exponent of the form
\begin{equation}
\label{exponent}
\begin{split}
&{- 4 \pi \tau_2 \over k} c^2 +  i (\beta + \bar{\beta}) c + 2 \pi i \bar{\tau} ({(j + k \bar{n} + {1 \over 2})^2 \over k} + \bar{\ell}) - \bar{\rho} (j + k \bar{n} + {1 \over 2}+\bar{m_2}) - \bar{\beta} \bar{m_1}\\
&- \rho(j + k n  + {1 \over 2} + m_2) - \beta m_1 - 2 \pi i {\tau} ({(j + k n + {1 \over 2})^2 \over k} + \ell )
\end{split}
\end{equation}
Our notation is slightly different from \cite{Maldacena:2000kv}. The terms $\ell, \bar{{\ell}}, \bar{m_1} , m_1, \bar{m_2}, m2$ merely come from expanding out all the terms in the partition function \eqref{susypart} and we will consider them in more detail in a moment. 

The integral over $\tau_2$ splits up into winding sectors, with the winding sector $w$ spanning the 
range ${b \over 2 \pi w} < \tau_2 < {b \over 2 \pi (w + 1)}$, where as usual $b = {\rm Re}(\beta)$. 
The integral over $c$ picks up poles at:
\begin{equation}
{-c^2 \over k} = {(j + k \bar{n} + {1 \over 2})^2 \over k} + \ell .
\end{equation}
In each winding sector, we have the constraint,
\begin{equation}
\label{cbounds}
{k w \over 2} < {\rm Im} c < {k (w + 1) \over 2} ,
\end{equation}
while the integral over $\tau_1$ yields the level matching condition
\begin{equation}
\label{levelmatching}
{(j + k n + {1 \over 2})^2 \over k} + \ell = {(j + k \bar{n} + {1 \over 2})^2 \over k} + \bar{\ell} .
\end{equation}

Consider the anti-holomorphic part of equation \eqref{exponent}. Doing the integral over $c$ yields the term:
\begin{equation}
-\bar{\beta} \left(\bar{m_1} + \sqrt{k({(j + k \bar{n} + {1 \over 2})^2 \over k} + \bar{\ell})}\right) - \bar{\rho}(\bar{m_2} + j + k \bar{n} + {1 \over 2}) 
\end{equation}
Now, note that for this term to survive in the limit $\bar{\beta} = -\bar{\rho} + \bar{\mu} \rightarrow \infty$ , we need to have:
\begin{equation}
\label{chiralprimcondition}
\bar{m_2} + j + k \bar{n} + {1 \over 2} = \bar{m_1} + \sqrt{k({(j + k \bar{n} + {1 \over 2})^2 \over k} + \bar{\ell})} .
\end{equation}
We will now show that this can happen, only if in the expansion of the partition function, we include only `zero-modes' and no `oscillator modes'. To lighten the notation, define
\begin{equation}
\bar{t} = j + k \bar{n} + {1 \over 2} ,~~ \delta = \bar{m_2} - \bar{m_1} .
\end{equation}
If $\delta \neq 0$ then equation \eqref{chiralprimcondition} has a solution subject to the constraints \eqref{cbounds} when
\begin{equation}
\label{solutioncondition}
{k w \over 2} < \delta + t = {k \bar{\ell} \over 2 \delta} + {\delta \over 2} < k {w + 1 \over 2} .
\end{equation}
This inequality implies $\delta > 0$ and we will show, that in this case, 
\begin{equation}
\label{chiralinequality}
\bar{\ell}  \geq \delta (w + 1) .
\end{equation}
Hence, the a solution to \eqref{chiralprimcondition} can never be found, except at $\bar{\ell} = 0, \delta = 0$. 

Let us write out some of the $\theta$ functions in \eqref{susypart} explicitly:
\begin{equation}
\label{thetaexplicit}
\begin{split}
&{\theta_2(i {{\bar{\beta}} - {\bar{\rho}} \over 4 \pi}, \bar{q})^2 \over \theta_1({i {\bar{\beta}} \over 2 \pi}, \bar{q}) \theta_1({i {\bar{\rho}} \over 2 \pi}, \bar{q})} = {(1 + e^{{{\bar{\rho}} - {\bar{\beta}} \over 2}})^2 \over (1 - e^{-{\bar{\beta}}})(1 - e^{{\bar{\rho}}})} \prod_{n=1}^{\infty}{(1 + {\bar{q}}^n e^{-{{\bar{\beta}} - {\bar{\rho}} \over 2}})^2 (1 + {\bar{q}}^n e^{+{{\bar{\beta}} - {\bar{\rho}} \over 2}})^2 \over (1 - {\bar{q}}^n e^{-{\bar{\beta}}}) (1 - {\bar{q}}^n e^{{\bar{\beta}}}) (1 - {\bar{q}}^n e^{-{\bar{\rho}}}) (1 - {\bar{q}}^n e^{{\bar{\rho}}})} \\
&= (-1)^w \prod_{n=0}^{w} {(e^{-{{\bar{\beta}} - {\bar{\rho}} \over 2}} {\bar{q}}^{-n} + 1)^2 \over (e^{-{\bar{\beta}}} {\bar{q}}^{-n} - 1)(e^{{\bar{\rho}}} {\bar{q}}^n - 1)} \prod_{n=w+1}^{\infty}{(1 + {\bar{q}}^n e^{ {{\bar{\beta}} - {\bar{\rho}} \over 2}})^2 \over (1 - {\bar{q}}^n e^{{\bar{\beta}}}) (1 - {\bar{q}}^n e^{-{\bar{\rho}}})} \\ &\times \prod_{n=1}^{\infty}{(1 + {\bar{q}}^n e^{-{{\bar{\beta}} - {\bar{\rho}} \over 2}})^2 \over (1 - {\bar{q}}^n e^{-{\bar{\beta}}}) (1 - {\bar{q}}^n e^{{\bar{\rho}}})} .
\end{split}
\end{equation}
The reason we transformed the first line above into the second line is for ease in series expansion. The integral \eqref{susypart} has poles when $\tau_2 = {b \over 2 \pi w}$ so one has to be careful while expanding in powers of $e^{-{\bar{\beta}}}$. Here, we are in the regime where, ${b \over 2 \pi(w+1)} < \tau_2 < {b \over 2 \pi w}$. So, in the second line above, we can expand all terms of the form ${1 \over 1 - x}$ as $\sum_{0}^{\infty} x^n$. 
Now, notice that for each term, \eqref{chiralinequality} holds. The first product which goes from $1 \ldots w$ has ${\bar m}_2 < 0, {\bar m}_1 > 0, {\bar{\ell}} < 0$ but $|{\bar m}_1 - {\bar m}_2| \geq {|{\bar{\ell}}| \over w+1}$. The second product which goes from $w+1 \ldots \infty$, has ${\bar m}_2 > 0, {\bar m}_1 < 0, {\bar{\ell}} > 0$ but ${\bar m}_2 - {\bar m}_1 \leq {{\bar{\ell}} \over w+1}$. The third product has ${\bar m}_1 < 0, {\bar m}_2 < 0, {\bar{\ell}} > 0$, so it also satisfies \eqref{chiralinequality}. The other important term in \eqref{susypart} is $\theta_2(i {{\bar{\beta}} + {\bar{\rho}} \over 4 \pi}, \bar{q})^2$. Every term in the expansion of this theta function has $\delta = 0$. Hence, the only terms that can satisfy \eqref{chiralprimcondition} are the {\em zero} modes of this theta function that also have ${\bar{\ell}} = 0$.
 It is apparent that \eqref{chiralinequality} holds for the eta functions in \eqref{susypart}. To conclude, we need to consider only the zero-modes in $\theta_2(i {{\bar{\beta}} + {\bar{\rho}} \over 4 \pi})$ and we can neglect everything else in the limit ${\bar{\beta}} = -{\bar{\rho}} + {\bar{\mu}} \rightarrow \infty$.

A very similar argument works for the contribution from the continuous representations. The contribution of the continuous representations comes from the divergences in the integral \eqref{susypart} near $\tau_2 = {b \over 2 \pi w}$. To analyze these, we replace $\tau$ by its value at the pole everywhere except in the divergent term and then  again expand out the partition function. By the argument above, again, we only need to concern ourselves with zero modes. In the limit $\bar{\beta} = - \bar{\rho} \rightarrow \infty$, the contribution from this pole vanishes unless: 
 \begin{equation}
j + k \bar{n} + {1 \over 2} = {k w \over 4} + {1 \over w} ({s^2 \over k} + {(j + k \bar{n} + {1 \over 2})^2 \over k}) 
\end{equation}
is met. This can only happen if:
\begin{equation}
\begin{split}
s &= 0, \\
j + k \bar{n} + {1 \over 2} &= {k w \over 2} .
\end{split}
\end{equation}
However, this condition can never be met because the sum over $j$ runs from $0 \ldots {k \over 2} - 1$. Thus it is precisely the chiral primaries that would have been in the continuum that are missing from our list above

\subsubsection{${1 \over 4}$ BPS partition function}
${1 \over 4}$ BPS states are of the form $|{\rm anything} \rangle |{\rm chiral~primary}\rangle$. The first step is to extract the anti-holomorphic chiral 
primary from the integral, as detailed above. Then, we merely need to series
expand the holomorphic term and pick out the term that satisfied the level matching
condition \eqref{levelmatching}. They key property we need here is
\begin{equation}
\begin{split}
{\theta_2({u - v \over 2} + w \tau)^2 \theta_2 ({u + v \over 2})^2  \over -i \theta_1(u + w \tau) \eta(\tau)^6} & \chi_{j}^{Q_5-2}(v - w \tau)\\ &= \left\{ \begin{array}{l l}  z^{Q_5 w \over 2} q^{-{Q_5 w^2 \over 4}} {\theta_2({u - v \over 2})^2 \theta_2 ({u + v \over 2})^2 \over -i \theta_1(u ) \eta(\tau)^6} \chi_{j}^{Q_5-2}(v) & \mbox{w even}; \\
 - z^{Q_5 w \over 2} q^{- {Q_5 w^2 \over 4}} {\theta_2({u - v \over 2})^2 \theta_2 ({u + v \over 2})^2 \over -i \theta_1(u) \eta(\tau)^6} \chi_{{Q_5 \over 2} - j - 1}^{Q_5-2}(v) & \mbox{w odd}, \end{array} \right.
\end{split}
\end{equation}
where as usual $z = e^{2 \pi i v}, q = e^{2 \pi i \tau}$. We can use this to shift the arguments of the $\theta$ function to a regime where \eqref{fdef} is applicable. Then \eqref{quarterbpspart} follows.

\subsubsection{Elliptic Genus}
To obtain the elliptic genus, we should take $\bar{\rho} = -\bar{\beta} + 2 \pi i$. As we mentioned, the partition function \eqref{susypart} vanishes with this substitution due to the zero mode contributions from the $\theta$ functions in the numerator. 
Evaluating the modified index \eqref{modifiedindex} is equivalent 
to replacing this term with a constant, which in our normalization is $-{1 \over 2}$.  Apart from this we see that with 
these chemical potentials, dramatic cancellations occur in formula \eqref{susypart}. We find
\begin{equation}
\label{ellipticappendix}
\begin{split}
&E_2(\beta, \rho) \sim {-b Q_5^{1 \over 2} \over 2 \pi} \int {d \tau_2  \over \tau_2^{3 \over 2}} \int_{-1 \over 2}^{1 \over 2} d \tau_1  e^{- Q_5 b^2  \over 4 \pi \tau_2} \left({1  \over \theta_1({i \beta \over 2 \pi}, \tau)} \right)  \\
&\times {4 \theta_2(i {\beta + \rho \over 4 \pi})^2 \theta_2(i {\beta - \rho \over 4 \pi})^2 \over \eta(\tau)^6} \times \sum_{j = 0}^{{Q_5 \over 2} - 1} \left( {\sum_{n \in \mathbb{Z}}  q^{(j + {1 \over 2} + Q_5 n)^2/Q_5}(z^{j + {1 \over 2} + Q_5 n} - z^{-(j + {1 \over 2} + Q_5 n)}) \over \theta_1({i \rho \over 2 \pi}, \tau)} \right.  \\
 &\times \left. (-{1 \over 2}) \cdot {(1 + e^{-\bar{\beta} + \bar{\rho} \over 2})^2 {\sum_{m \in \mathbb{Z}}\bar{q}^{(j + {1 \over 2} + Q_5 m)^2/Q_5}(\bar{z}^{j + {1 \over 2} + Q_5 m} - \bar{z}^{-(j + {1 \over 2} + Q_5 m)}) \over (1 - e^{-\bar{\beta}})(1 - e^{\bar{\rho}})} }\right) .
\end{split}
\end{equation}
In the last line, we will interpret the zero modes of the theta functions that 
appear as the action of the global generators of the {\em spacetime} supergroup
$SU(1,1|2)$.  The term $(1 - e^{-\bar{\beta}})$ which corresponds to an operator with
$\bar{h} = 1, \bar{r} = 0$ represents the action of $\bar{L}^b_{-1}$. The term $(1 - e^{\bar{\rho}})$ represents an operator with $\bar{h} = 0, \bar{r} = -1$ and corresponds to the action of $\bar{K}^{-}$ the lowering operator of the $SU(2)$ R-symmetry. $\bar{K}^{-}$ acts on the term in the numerator $(\bar{z}^{j + {1 \over 2} + k m}  - \bar{z}^{-(j + {1 \over 2} + k m)})$ to generate a $SU(2)$ representation. The two other terms in the numerator $(1 + e^{-\bar{\beta} + \bar{\rho} \over 2})^2$ correspond to fermionic operators with $\bar{h} = {1 \over 2}, \bar{r} = -{1 \over 2}$. This term represents
the action of the two global supercharges that do not annihilate the chiral primary at the head of this representation.

If $m > 0$, this chiral primary is represented by the term $\bar{z}^{j + {1 \over 2} + Q_5 m}$ in the $SU(2)$ representation. In this case, the term $\bar{z}^{-(j + {1 \over 2} + Q_5 m)}$ represents
an anti-chiral primary (The reverse is true for $m < 0$).  The surviving global supercharges should annihilate this term. 
 It appears that in the formula \eqref{susypart} we need to impose this projection by hand. This is equivalent to dropping the term $z^{-(j+ {1 \over 2} + k m)}$ in \eqref{ellipticappendix} 
and leads to formula \eqref{ellipticsimple}. The same projection needs to be imposed on the holomorphic term in \eqref{ellipticsimple} and \eqref{susypart}. This deserves a better understanding.

\section{Theta Functions}
\label{thetaappendix}
Here we list our convention for various theta functions.
We define:
\begin{equation}
\label{thetadef}
\theta(a,b)(\nu, \tau) = \sum_{p \in \mathbb{Z}} e^{ \pi i \tau (p + {a \over 2})^2 + 2 \pi i (\nu + {b \over 2})(p + {a \over 2})} ,
\end{equation}
with the conventions:
\begin{equation}
\begin{split}
\theta_1 &= \theta(1,1) ,\\
\theta_2 &= \theta(1,0) , \\
\theta_3 &=  \theta(0,0) , \\ 
\theta_4 &= \theta(0,1) .
\end{split}
\end{equation}

Defining, $q = e^{2\pi i \tau}$ and $z = e^{2 \pi i \rho}$ the definitions above lead to the following product formulae \cite{DiFrancesco:1997nk}.
\begin{equation}
\begin{split}
\theta_1(\rho, \tau) &= -i z^{1 \over 2} q^{1 \over 8} \prod_{n=1}^{\infty} (1 - q^n) \prod_{n=0}^{\infty}(1 - z q^{n+1})(1 - z^{-1} q^n) ,\\
\theta_2(\rho,\tau)&=  z^{1 \over 2} q^{1 \over 8} \prod_{n=1}^{\infty} (1 - q^{n}) \prod_{n=0}^{\infty}(1 + z q^{n+1})(1 + z^{-1} q^n) ,\\
\theta_3(\rho,\tau) &= \prod_{n=1}^{\infty} (1 - q^n) \prod_{r \in \mathbb{N}+ 1/2} (1 + z q^r)(1 + z^{-1} q^r) ,\\
\theta_4(\rho,\tau) &= \prod_{n=1}^{\infty} (1 - q^n) \prod_{r \in \mathbb{N}+ 1/2} (1 - z q^r)(1 - z^{-1} q^r) .\\
\end{split}
\end{equation}
We sometimes use the abbreviated notation $\theta(\rho)$ for $\theta(\rho, \tau)$.
The $\eta$ function is defined by:
\begin{equation} 
\eta(\tau) = q^{1 \over 24} \prod_{n=1}^{\infty}(1 - q^n) .
\end{equation}

\bibliographystyle{JHEP}
\bibliography{references}

\providecommand{\href}[2]{#2}\begingroup\raggedright\begin{thebibliography}{10}

\bibitem{Mandal}
G.~Mandal, S.~Raju, and M.~Smedback, {\it {Supersymmetric Giant Graviton
  Solutions in AdS(3)}},  \href{http://xxx.lanl.gov/abs/{arXiv:0709.1168
  [hep-th]}}{{\tt {arXiv:0709.1168 [hep-th]}}}.

\bibitem{Maldacena:1997re}
J.~M. Maldacena, {\it {The large N limit of superconformal field theories and
  supergravity}},  {\em Adv. Theor. Math. Phys.} {\bf 2} (1998) 231--252,
  [\href{http://xxx.lanl.gov/abs/hep-th/9711200}{{\tt hep-th/9711200}}].

\bibitem{Seiberg:1999xz}
N.~Seiberg and E.~Witten, {\it {The D1/D5 system and singular CFT}},  {\em
  JHEP} {\bf 04} (1999) 017,
  [\href{http://xxx.lanl.gov/abs/hep-th/9903224}{{\tt hep-th/9903224}}].

\bibitem{Kinney:2005ej}
J.~Kinney, J.~M. Maldacena, S.~Minwalla, and S.~Raju, {\it An index for 4
  dimensional super conformal theories},  {\em Commun. Math. Phys.} {\bf 275}
  (2007) 209--254, [\href{http://xxx.lanl.gov/abs/hep-th/0510251}{{\tt
  hep-th/0510251}}].

\bibitem{Maldacena:1998bw}
J.~M. Maldacena and A.~Strominger, {\it {AdS(3) black holes and a stringy
  exclusion principle}},  {\em JHEP} {\bf 12} (1998) 005,
  [\href{http://xxx.lanl.gov/abs/hep-th/9804085}{{\tt hep-th/9804085}}].

\bibitem{deBoer:1998us}
J.~de~Boer, {\it {Large N Elliptic Genus and AdS/CFT Correspondence}},  {\em
  JHEP} {\bf 05} (1999) 017,
  [\href{http://xxx.lanl.gov/abs/hep-th/9812240}{{\tt hep-th/9812240}}].

\bibitem{deBoer:1998ip}
J.~de~Boer, {\it {Six-dimensional supergravity on S**3 x AdS(3) and 2d
  conformal field theory}},  {\em Nucl. Phys.} {\bf B548} (1999) 139--166,
  [\href{http://xxx.lanl.gov/abs/hep-th/9806104}{{\tt hep-th/9806104}}].

\bibitem{Dijkgraaf:2000fq}
R.~Dijkgraaf, J.~M. Maldacena, G.~W. Moore, and E.~P. Verlinde, {\it {A black
  hole farey tail}},  \href{http://xxx.lanl.gov/abs/hep-th/0005003}{{\tt
  hep-th/0005003}}.

\bibitem{Maldacena:2000hw}
J.~M. Maldacena and H.~Ooguri, {\it {Strings in AdS(3) and SL(2,R) WZW model.
  I}},  {\em J. Math. Phys.} {\bf 42} (2001) 2929--2960,
  [\href{http://xxx.lanl.gov/abs/hep-th/0001053}{{\tt hep-th/0001053}}].

\bibitem{Maldacena:2000kv}
J.~M. Maldacena, H.~Ooguri, and J.~Son, {\it {Strings in {AdS(3)} and the
  SL(2,R) WZW model. II: Euclidean black hole}},  {\em J. Math. Phys.} {\bf 42}
  (2001) 2961--2977, [\href{http://xxx.lanl.gov/abs/hep-th/0005183}{{\tt
  hep-th/0005183}}].

\bibitem{Grant:2005qc}
L.~Grant, L.~Maoz, J.~Marsano, K.~Papadodimas, and V.~S. Rychkov, {\it
  {Minisuperspace quantization of 'bubbling AdS' and free fermion droplets}},
  {\em JHEP} {\bf 08} (2005) 025,
  [\href{http://xxx.lanl.gov/abs/hep-th/0505079}{{\tt hep-th/0505079}}].

\bibitem{Mandal:2005wv}
G.~Mandal, {\it {Fermions from half-BPS supergravity}},  {\em JHEP} {\bf 08}
  (2005) 052, [\href{http://xxx.lanl.gov/abs/hep-th/0502104}{{\tt
  hep-th/0502104}}].

\bibitem{Maoz:2005nk}
L.~Maoz and V.~S. Rychkov, {\it {Geometry quantization from supergravity: The
  case of 'bubbling AdS'}},  {\em JHEP} {\bf 08} (2005) 096,
  [\href{http://xxx.lanl.gov/abs/hep-th/0508059}{{\tt hep-th/0508059}}].

\bibitem{Rychkov:2005ji}
V.~S. Rychkov, {\it {D1-D5 black hole microstate counting from supergravity}},
  {\em JHEP} {\bf 01} (2006) 063,
  [\href{http://xxx.lanl.gov/abs/hep-th/0512053}{{\tt hep-th/0512053}}].

\bibitem{Biswas:2006tj}
I.~Biswas, D.~Gaiotto, S.~Lahiri, and S.~Minwalla, {\it {Supersymmetric states
  of N = 4 Yang-Mills from giant gravitons}},
  \href{http://xxx.lanl.gov/abs/hep-th/0606087}{{\tt hep-th/0606087}}.

\bibitem{Mandal:2006tk}
G.~Mandal and N.~V. Suryanarayana, {\it {Counting 1/8-BPS dual-giants}},  {\em
  JHEP} {\bf 03} (2007) 031,
  [\href{http://xxx.lanl.gov/abs/hep-th/0606088}{{\tt hep-th/0606088}}].

\bibitem{Martelli:2006vh}
D.~Martelli and J.~Sparks, {\it {Dual giant gravitons in Sasaki-Einstein
  backgrounds}},  {\em Nucl. Phys.} {\bf B759} (2006) 292--319,
  [\href{http://xxx.lanl.gov/abs/hep-th/0608060}{{\tt hep-th/0608060}}].

\bibitem{Basu:2006id}
A.~Basu and G.~Mandal, {\it {Dual giant gravitons in AdS(m) x Y**n
  (Sasaki-Einstein)}},  \href{http://xxx.lanl.gov/abs/hep-th/0608093}{{\tt
  hep-th/0608093}}.

\bibitem{Argurio:2000tb}
R.~Argurio, A.~Giveon, and A.~Shomer, {\it {Superstrings on AdS(3) and
  symmetric products}},  {\em JHEP} {\bf 12} (2000) 003,
  [\href{http://xxx.lanl.gov/abs/hep-th/0009242}{{\tt hep-th/0009242}}].

\bibitem{Dijkgraaf:1998gf}
R.~Dijkgraaf, {\it {Instanton strings and hyperKaehler geometry}},  {\em Nucl.
  Phys.} {\bf B543} (1999) 545--571,
  [\href{http://xxx.lanl.gov/abs/hep-th/9810210}{{\tt hep-th/9810210}}].

\bibitem{dedecker1953cvf}
P.~Dedecker, {\it {Calcul des variations, formes diff{\'e}rentielles et champs
  geodesiques}},  {\em Geometrie diff{\'e}rentielle, Colloq. Intern. du CNRS
  LII, Strasbourg} (1953) 17--34.

\bibitem{goldschmidt1973hcf}
H.~Goldschmidt and S.~Sternberg, {\it {The Hamilton-Cartan formalism in the
  calculus of variations}},  {\em Ann. Inst. Fourier} {\bf 23} (1973), no.~1
  203--267.

\bibitem{Kijowski:1973gi}
J.~Kijowski, {\it {A finite-dimensional canonical formalism in the classical
  field theory}},  {\em Commun. Math. Phys.} {\bf 30} (1973) 99--128.

\bibitem{gawedzki1974cfl}
K.~Gawedzki and W.~Kondracki, {\it {Canonical formalism for the local-type
  functionals in the classical field theory}},  {\em Rep. Math. Phys} {\bf 6}
  (1974) 465--476.

\bibitem{szczyrba1976sss}
W.~Szczyrba, {\it {A symplectic structure on the set of Einstein metrics}},
  {\em Communications in Mathematical Physics} {\bf 51} (1976), no.~2 163--182.

\bibitem{garcia570rss}
P.~Garcia, {\it {Reducibility of the symplectic structure of classical fields
  with gauge symmetry}},  {\em Lecture Notes in Mathematics} {\bf 570} (1977).

\bibitem{zuckerman1987apa}
G.~Zuckerman, {\it {Action principles and global geometry}},  {\em Mathematical
  Aspects of String Theory} {\bf 1} (1987) 259--284.

\bibitem{crnkovic1987cdc}
C.~Crnkovic and E.~Witten, {\it {Covariant description of canonical formalism
  in geometrical theories}},  {\em Three Hundred Years of Gravitation} (1987)
  676--684.

\bibitem{Lee:1990nz}
J.~Lee and R.~M. Wald, {\it {Local symmetries and constraints}},  {\em J. Math.
  Phys.} {\bf 31} (1990) 725--743.

\bibitem{Witten:1983ar}
E.~Witten, {\it {Nonabelian bosonization in two dimensions}},  {\em Commun.
  Math. Phys.} {\bf 92} (1984) 455--472.

\bibitem{Giveon:1998ns}
A.~Giveon, D.~Kutasov, and N.~Seiberg, {\it {Comments on string theory on
  AdS(3)}},  {\em Adv. Theor. Math. Phys.} {\bf 2} (1998) 733--780,
  [\href{http://xxx.lanl.gov/abs/hep-th/9806194}{{\tt hep-th/9806194}}].

\bibitem{Chu:1991pn}
M.-f. Chu, P.~Goddard, I.~Halliday, D.~I. Olive, and A.~Schwimmer, {\it
  {Quantization of the Wess-Zumino-Witten model on a circle}},  {\em Phys.
  Lett.} {\bf B266} (1991) 71--81.

\bibitem{Lunin:2002bj}
O.~Lunin, S.~D. Mathur, and A.~Saxena, {\it {What is the gravity dual of a
  chiral primary?}},  {\em Nucl. Phys.} {\bf B655} (2003) 185--217,
  [\href{http://xxx.lanl.gov/abs/hep-th/0211292}{{\tt hep-th/0211292}}].

\bibitem{Gaberdiel:2007vu}
M.~R. Gaberdiel and I.~Kirsch, {\it {Worldsheet correlators in AdS(3)/CFT(2)}},
   {\em JHEP} {\bf 04} (2007) 050,
  [\href{http://xxx.lanl.gov/abs/hep-th/0703001}{{\tt hep-th/0703001}}].

\bibitem{Dabholkar:2007ey}
A.~Dabholkar and A.~Pakman, {\it {Exact chiral ring of AdS(3)/CFT(2)}},
  \href{http://xxx.lanl.gov/abs/hep-th/0703022}{{\tt hep-th/0703022}}.

\bibitem{DiFrancesco:1997nk}
P.~Di~Francesco, P.~Mathieu, and D.~Senechal, {\it {Conformal field theory}}, .
  New York, USA: Springer (1997) 890.

\bibitem{Dixon:1989cg}
L.~J. Dixon, M.~E. Peskin, and J.~D. Lykken, {\it {N=2 Superconformal Symmetry
  and SO(2,1) Current Algebra}},  {\em Nucl. Phys.} {\bf B325} (1989) 329--355.

\bibitem{Henningson:1991jc}
M.~Henningson, S.~Hwang, P.~Roberts, and B.~Sundborg, {\it {Modular invariance
  of SU(1,1) strings}},  {\em Phys. Lett.} {\bf B267} (1991) 350--355.

\bibitem{Felder:1988as}
G.~Felder, K.~Gawedzki, and A.~Kupiainen, {\it {The Spectrum of
  Wess-Zumino-Witten Models}},  {\em Nucl. Phys.} {\bf B299} (1988) 355--366.

\bibitem{Gawedzki:1990jc}
K.~Gawedzki, {\it {Classical origin of quantum group symmetries in
  Wess-Zumino-Witten conformal field theory}},  {\em Commun. Math. Phys.} {\bf
  139} (1991) 201--214.

\bibitem{Barabanschikov:2005ri}
A.~Barabanschikov, L.~Grant, L.~L. Huang, and S.~Raju, {\it {The spectrum of
  Yang Mills on a sphere}},  {\em JHEP} {\bf 01} (2006) 160,
  [\href{http://xxx.lanl.gov/abs/hep-th/0501063}{{\tt hep-th/0501063}}].

\bibitem{Callan:1991at}
J.~Callan, Curtis~G., J.~A. Harvey, and A.~Strominger, {\it {Supersymmetric
  string solitons}},  \href{http://xxx.lanl.gov/abs/hep-th/9112030}{{\tt
  hep-th/9112030}}.

\bibitem{Callan:1991dj}
J.~Callan, Curtis~G., J.~A. Harvey, and A.~Strominger, {\it {World sheet
  approach to heterotic instantons and solitons}},  {\em Nucl. Phys.} {\bf
  B359} (1991) 611--634.

\bibitem{Callan:1991ky}
J.~Callan, Curtis~G., J.~A. Harvey, and A.~Strominger, {\it {Worldbrane actions
  for string solitons}},  {\em Nucl. Phys.} {\bf B367} (1991) 60--82.

\bibitem{Kutasov:1999xu}
D.~Kutasov and N.~Seiberg, {\it {More comments on string theory on AdS(3)}},
  {\em JHEP} {\bf 04} (1999) 008,
  [\href{http://xxx.lanl.gov/abs/hep-th/9903219}{{\tt hep-th/9903219}}].

\bibitem{deBoer:1998pp}
J.~de~Boer, H.~Ooguri, H.~Robins, and J.~Tannenhauser, {\it {String theory on
  AdS(3)}},  {\em JHEP} {\bf 12} (1998) 026,
  [\href{http://xxx.lanl.gov/abs/hep-th/9812046}{{\tt hep-th/9812046}}].

\bibitem{Giveon:2003ku}
A.~Giveon and A.~Pakman, {\it {More on superstrings in AdS(3) x N}},  {\em
  JHEP} {\bf 03} (2003) 056,
  [\href{http://xxx.lanl.gov/abs/hep-th/0302217}{{\tt hep-th/0302217}}].

\bibitem{Pakman:2003cu}
A.~Pakman, {\it {Unitarity of supersymmetric SL(2,R)/U(1) and no-ghost theorem
  for fermionic strings in AdS(3) x N}},  {\em JHEP} {\bf 01} (2003) 077,
  [\href{http://xxx.lanl.gov/abs/hep-th/0301110}{{\tt hep-th/0301110}}].

\bibitem{Pakman:2003kh}
A.~Pakman, {\it {BRST quantization of string theory in AdS(3)}},  {\em JHEP}
  {\bf 06} (2003) 053, [\href{http://xxx.lanl.gov/abs/hep-th/0304230}{{\tt
  hep-th/0304230}}].

\bibitem{Israel:2003ry}
D.~Israel, C.~Kounnas, and M.~P. Petropoulos, {\it {Superstrings on NS5
  backgrounds, deformed AdS(3) and holography}},  {\em JHEP} {\bf 10} (2003)
  028, [\href{http://xxx.lanl.gov/abs/hep-th/0306053}{{\tt hep-th/0306053}}].

\bibitem{Aharony:1999ti}
O.~Aharony, S.~S. Gubser, J.~M. Maldacena, H.~Ooguri, and Y.~Oz, {\it {Large N
  field theories, string theory and gravity}},  {\em Phys. Rept.} {\bf 323}
  (2000) 183--386, [\href{http://xxx.lanl.gov/abs/hep-th/9905111}{{\tt
  hep-th/9905111}}].

\bibitem{Maldacena:1999bp}
J.~M. Maldacena, G.~W. Moore, and A.~Strominger, {\it {Counting BPS black holes
  in toroidal type II string theory}},
  \href{http://xxx.lanl.gov/abs/hep-th/9903163}{{\tt hep-th/9903163}}.

\bibitem{Kutasov:1998zh}
D.~Kutasov, F.~Larsen, and R.~G. Leigh, {\it {String theory in magnetic
  monopole backgrounds}},  {\em Nucl. Phys.} {\bf B550} (1999) 183--213,
  [\href{http://xxx.lanl.gov/abs/hep-th/9812027}{{\tt hep-th/9812027}}].

\bibitem{Kraus:2006nb}
P.~Kraus and F.~Larsen, {\it {Partition functions and elliptic genera from
  supergravity}},  {\em JHEP} {\bf 01} (2007) 002,
  [\href{http://xxx.lanl.gov/abs/hep-th/0607138}{{\tt hep-th/0607138}}].

\bibitem{Cecotti:1992qh}
S.~Cecotti, P.~Fendley, K.~A. Intriligator, and C.~Vafa, {\it {A New
  supersymmetric index}},  {\em Nucl. Phys.} {\bf B386} (1992) 405--452,
  [\href{http://xxx.lanl.gov/abs/hep-th/9204102}{{\tt hep-th/9204102}}].

\bibitem{fuchs1997sla}
J.~Fuchs and C.~Schweigert, {\em {Symmetries, Lie algebras and
  representations}}.
\newblock Cambridge University Press New York, NY, USA, 1997.

\end{thebibliography}\endgroup

\end{document}